\documentclass{emulateapj}

\input{epsf}

\begin{document}

\title{A survey of the high order multiplicity of nearby solar-type binary stars with Robo-AO}

\author{Reed L. Riddle\altaffilmark{1}, Andrei Tokovinin\altaffilmark{2},
Brian D.\ Mason\altaffilmark{3}, William I.\ Hartkopf\altaffilmark{3},
Lewis C.\ Roberts, Jr.\altaffilmark{4}, Christoph Baranec\altaffilmark{5}, 
Nicholas M. Law\altaffilmark{6}, 
Khanh Bui\altaffilmark{1}, Mahesh P. Burse\altaffilmark{7}, H. K. Das\altaffilmark{7}, 
Richard G. Dekany\altaffilmark{1}, Shrinivas Kulkarni\altaffilmark{1}, 
Sujit Punnadi\altaffilmark{7}, A. N. Ramaprakash\altaffilmark{7}, Shriharsh P. Tendulkar\altaffilmark{1}
  }

\affil{\altaffilmark{1}Division of Physics, Mathematics, and Astronomy, California Institute of Technology, Pasadena, CA 91125, USA}
\affil{\altaffilmark{2}Cerro Tololo Inter-American Observatory, Casilla 603, La Serena, Chile}
\affil{\altaffilmark{3}U.S. Naval Observatory, 3450 Massachusetts Avenue, Washington, DC 20392-5420, USA}
\affil{\altaffilmark{4}Jet Propulsion Laboratory, California Institute of Technology, 4800 Oak Grove Drive, Pasadena, CA 91109, USA} 
\affil{\altaffilmark{5}Institute for Astronomy, University of Hawai$\textquoteleft$i at M\={a}noa, Hilo, HI 96720-2700, USA}
\affil{\altaffilmark{6}Department of Physics and Astronomy, University of North Carolina at Chapel Hill, Chapel Hill, NC 27599-3255, USA}
\affil{\altaffilmark{7}Inter-University Centre for Astronomy \& Astrophysics, Ganeshkhind, Pune, 411007, India}

\shorttitle{Multiplicity survey of nearby solar-type dwarfs with Robo-AO}
\shortauthors{Riddle et al.}
\slugcomment{To be submitted to ApJ}

\begin{abstract}

We conducted a survey of nearby binary systems composed of main sequence stars of spectral types F and G in order to improve our understanding of the hierarchical  nature of multiple star systems.  Using Robo-AO, the first robotic adaptive optics instrument, we collected high angular resolution images with deep and well-defined detection limits in the SDSS $i'$ band.  A total of  695 components belonging  to 595 systems were  observed.  We prioritized   observations   of   faint  secondary   components   with separations  over  $10''$ to  quantify  the  still poorly  constrained frequency of  their sub-systems.  Of the 214  secondaries observed, 39 contain such  subsystems; 19  of those were  discovered with  Robo-AO.  The selection-corrected  frequency of  secondary sub-systems  with periods from $10^{3.5}$ to $10^5$ days is 0.12$\pm$0.03, the same as the frequency of  such companions  to the  primary.  Half  of the secondary  pairs belong to  quadruple systems where the  primary is also  a close pair, showing that  the presence  of sub-systems in  both components  of the outer  binary is  correlated.   The relatively large  abundance of 2+2 quadruple systems is a new finding, and will require more exploration of the formation mechanism of multiple star systems.  We  also targeted  close  binaries with periods  less  than  100~yr,  searching for  their  distant  tertiary components, and discovered 17 certain and 2 potential new triples.  In a  sub-sample  of  241   close  binaries,  71  have  additional  outer companions.   The  overall frequency  of  tertiary  components is  not enhanced, compared  to all (non-binary)  targets, but in the  range of outer periods from $10^6$ to $10^{7.5}$ days (separations on the order of 500~AU),  the frequency  of tertiary components  is 0.16$\pm$0.03, exceeding by almost  a factor of two the  frequency of similar systems among all  targets (0.09). Measurements  of binary stars  with Robo-AO allowed us to  compute first orbits for 9 pairs  and to improve orbits of another 11 pairs.

\end{abstract}


\submitted{Draft version \today}

\section{Introduction}
\label{sec:intro}

Statistics of hierarchical stellar  systems provide important clues to
star  formation  mechanisms that  are  still  actively researched  and
debated.   Current  theories   give  contradictory  predictions  about
hierarchical multiples (for example, N-body dynamics vs. hydrodynamical 
simulations produce different multiplicity fractions);  so far, none  of them is capable  of modeling
the  distributions of periods,  mass ratios,  and hierarchies  seen in
observations.

As periods and separations of  stellar pairs vary by several orders of
magnitude,  comprehensive  coverage  of  the parameter  space  can  be
achieved  only by  combining complementary  observing  techniques, and
only  for the nearest  stars. However,  the number  of objects  in the
best-studied  25-pc sample  of \citet{Raghavan2010}  is too  small for
deriving meaningful  statistics of triple  and higher-order multiples.
For this  purpose, a larger, distance-limited sample  of $\sim$5000 F-
and G-dwarfs  within 67~pc of the  Sun was constructed  from the {\em
  Hipparcos}  catalog;  it  is  called FG-67  hereafter.   
  
The  sample
definition and the database are presented in \citet{Tokovinin2014a}.  
The census of companions to  primary stars in FG-67 is fairly complete
over the whole range of periods, with an overall detection probability
of  about 80\%,  except for  low-mass companions  at  separations from
0\farcs1   to  $20''$ (see details in the above paper).    Such   companions  can   be  discovered   by
high-contrast  imaging  with  adaptive  optics (AO).   Secondary
components can  be close pairs  as well, but discovery of  sub-systems in
the  {\em  secondary} components  is  more  problematic, with  average
detection probability  (before this survey)  of only about  15\%; most
sub-systems in the  secondaries are presently missed.  If  they are as
frequent as sub-systems  in the primaries, 
the  actual number  of  stellar  hierarchies in  the  FG-67 sample  is
substantially larger than known today.

It  is  important  to  test   the  frequency  of  sub-systems  in  the
secondaries  for constraining  mechanisms of  multiple-star formation.
If   chaotic    $N$-body   dynamics   is    the   dominant   mechanism, 
sub-systems in the secondaries should be
much  less common  than in  the primaries, because in triple  encounters, the
lightest star is usually ejected while the two most massive stars pair
in  a  binary.  On   the  other  hand,  the hydrodynamic  simulations  of
\citet{Bate2012} predict a large fraction of multiple systems, including
2+2 quadruples (i.e., two close pairs in a wide outer binary).

Recent  advances in   observational  techniques  have substantially
improved our knowledge of the  distribution of binary periods and mass
ratios,  as reviewed  by \citet{Duchene2013}.  Lucky imaging  has been
used      to     survey      large      samples     \citep{Ginski2012,
  Janson2012,Jodar2013}. Yet, the emphasis  is almost always placed on
binaries,  leaving higher-order hierarchies  aside or  addressing them
casually; a rare exception is the work of~\cite{2010ApJ...720.1727L} who observed a higher than expected high-order multiplicity of M dwarf stars.

This  survey  aims at  complementing  the  statistics of  hierarchical
systems  in  the FG-67  sample  by  imaging  with Robo-AO~\citep{Baranec2014}.   Our main
goal is to constrain the multiplicity of secondary components, most of
which so far had no high-resolution imaging data.  The second goal is
to search  for additional companions around primary  targets which are
themselves known  close binaries. Here  we explore the  poorly covered
part  of  the  parameter space  at  separations  on  the order  of  an
arcsecond, searching  for low-mass tertiary components.   In short, we
focus on binaries (both wide  and close) and quantify the frequency of
additional companions.

We  begin with  a brief  characterization  of the  surveyed sample  in
\S~\ref{sec:sample}.   Observations   with   Robo-AO   and   data
processing  are  covered in  \S~\ref{sec:obs},  with data  tables
described  in  \S~\ref{sec:data}.  In  \S~\ref{sec:res}  the
results of  this survey  are presented, and  \S~\ref{sec:sum}  is the
summary.

\section{The target sample}
\label{sec:sample}

The targets  for  this  survey   were  selected  from  the  FG-67  database
\citep{Tokovinin2014a}. We  observed known and  suspected binary stars
and tried to constrain the frequency of additional components in those
systems; high priority was  placed on distant secondary components in
wide  binaries.  The  selection  criteria were separations of $>10''$  and a
declination north of $-15^\circ$. Some brighter secondaries are in the
{\em Hipparcos} catalog, and  therefore ``screened'' for companions by
that space mission (but not as deeply as with Robo-AO).  The majority,
however, are rather faint.   Many of those secondaries were discovered
only recently by dedicated programs \citep{Tokovinin2011, LEP}.

The second-priority list contains mainly targets which are binaries with
periods  $P<100$~yr (separation  $\sim$0\farcs5 at  50~pc distance).
The rationale for restricting the  periods is related to dynamical
stability: the semi-major  axis of the outer orbit  should be at least
three times larger than that of the  inner orbit, and should match well the
discovery space of  Robo-AO. \citet{Harrington1992} quantified a value
of 3  for the  ratio of periastron of the outer binary to  apastron 
of the  inner binary  as  the critical  factor for  long-term
stability  (assuming equal  masses). The  program was  later
complemented with observations  of additional resolved visual binaries
with separations  from 0\farcs2 to  2$''$, and $P>100$~yr.   Here the
chances  of  discovering wider  tertiary  components  are small.   The
chances of resolving a secondary  into a hereto unknown close pair are
also  small.  However,  observing  those stars  (mostly resolved)  was
useful for  improving their orbits  and for confirming  some uncertain
visual  pairs which  were resolved  only once  in the  past  and never
confirmed since. The program included repeated observations of several
known binaries as calibrators for data quality control.

\begin{figure}[!ht]
\plotone{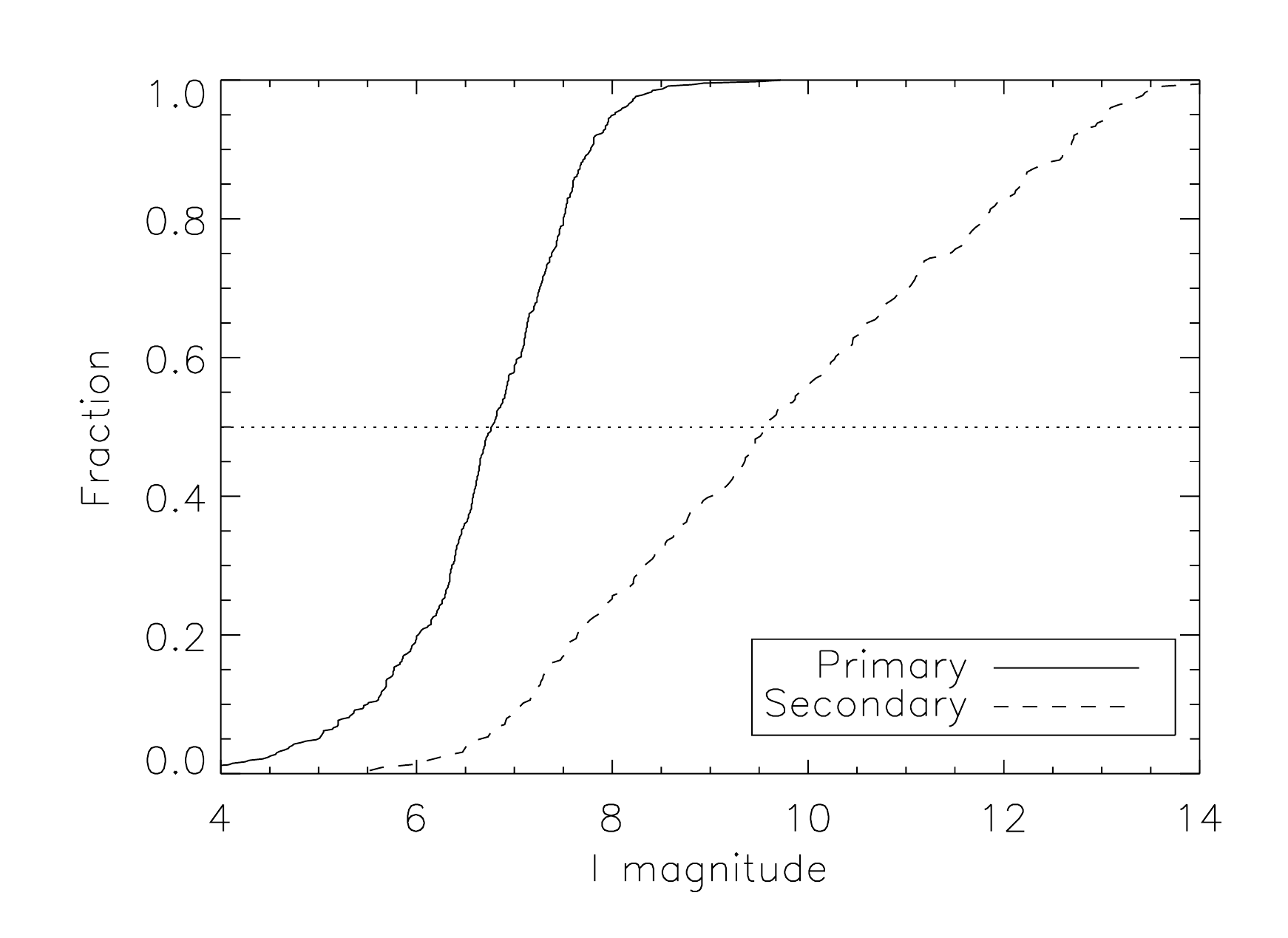}
\caption{Cumulative  histograms of  the $I_C$  magnitudes  of components observed with Robo-AO.
\label{fig:ihist} }
\end{figure}

We observed  most, but  not all, components  from the  original lists:  239 
secondaries (212 of those with good quality), 354 close pairs, and 102 resolved binaries.
Some secondaries  turned out to be  too faint.  Figure~\ref{fig:ihist}
presents  the distribution of  $I_C$ (Cousins  $I$) magnitudes  of our
primary targets  \citep{HIP2} and  of their secondary  components. The
latter  are derived  from their  $J$ magnitudes  in the  2MASS catalog
\citep{Cutri2003},   assuming  that   the  stars   are  on   the  main
sequence. All  secondaries with  useful data have  $I_C <  13.5$ mag.  The
median masses of primary and secondary components estimated from their
absolute magnitudes  are 1.22 and  0.71~${\cal M}_\odot$, respectively; 80\%
of secondary masses are between 0.31 and 1.04~${\cal M}_\odot$.

\section{Observations and data reduction}
\label{sec:obs}

The instrument  used for  this survey was  Robo-AO, the  first robotic
laser guide  star AO system \citep{Baranec2014}, which  is designed to
operate automatically  on 1-3~m class telescopes in  order to undertake
high efficiency  observing programs (e.g., large  surveys).  The prototype Robo-AO instrument is
currently deployed on the Palomar Observatory 60~inch (1.5~m) telescope (P60).

As  a fully  automated system,  Robo-AO is  unique in  its  ability to
observe  targets at both  high resolution  and high  cadence.  Robo-AO
uses Rayleigh  scattering from a UV  laser focused at  10~km from the
telescope  as the  wavefront reference,  and generates  images  at the
diffraction limit of the  P60 (0\farcs12 -- 0\farcs15 FWHM) with Strehl
ratios of 10-25\% in the visible filters used for this survey.  The AO
system corrects  the high  order wavefront aberrations  with automated
software  that  operates  at   a  rate  of  1.2~kHz,  sharpening  the
instantaneous point  spread function  (PSF) across the  science camera
field of view  (FOV).  A bright star within the  FOV is still required
to correct the tip-tilt motion; this is achieved in the automated data
processing  software  (\S~\ref{sec:lucky}).   Typically,  Robo-AO
requires a star brighter than $m \approx 16$ mag with a broad-band filter to apply the tip-tilt correction successfully.
Instrument    parameters    for   this    survey    are   listed    in
Table~\ref{tab:roboao}.

\subsection{Observations}
\label{sec:log}

\cite{Baranec2013} and \cite{Baranec2014} describe the operation and the science instrumentation of the Robo-AO system.  The output of the science camera raw data are image cubes  composed of 1024$\times$1024 pixel image frames; a total  of  516  frames were  gathered  during each  60~s
exposure, which  were then  combined into a  single image  for further
analysis  by  the  automated  data processing  software.   Almost  all
targets were observed with  the SDSS $i'$ filter \citep{York2000} with
a central    wavelength  of 754~nm  and    a FWHM   bandpass  of  119~nm.   
Some fainter targets  were observed  with a
long  pass  600~nm  filter  (LP600)  which  transmits  all
wavelengths longer than 600~nm to the quantum efficiency cutoff of the detector.  

The pointing error of the P60 telescope with Robo-AO mounted is on the
order of 10\arcsec,  and can vary between observing runs  depending on telescope
balance and adjustments.   With a 44\arcsec ~FOV, and  no clear way to
select one  target over another  in the stellar field,  Robo-AO does not
attempt to recenter the target  automatically in fear of selecting the
wrong  star and  moving the  science target  out of  the  field.  This
effect causes a large variation  in the placement of the target star
in  the image frame;  a planned  upgrade to  the Robo-AO  software and
continuing  improvements to the  P60 will  minimize this  effect.  All  observations were checked to confirm that the science target was captured in the CCD frame.  The
consequence for  this survey  is that target  placement is  across the
entire FOV, and  targets sometimes ended up uncomfortably close to the edge
of the  frame, limiting the  observable area around the  target.  When
necessary, stars affected were re-observed  to capture a larger area and
allow a better examination for companions.

Robo-AO can  observe targets at a  rate of $\sim$20 per  hour, with an
intelligent queue  system selecting  the best object  to observe  at a
given time  \citep{Riddle2014}.  The Robo-AO  queue system interleaves
several    different    science     programs    through    a    single
night. Observations of  this survey were spread over  a year (July 2012 to August 2013) to gather
targets at all hours of right ascension.  The  entire  695  targets  of this  survey  used
29 total hours of observing time (this time also included calibration binary observations and re-observations).  Robo-AO is ideally suited to
a large survey  such as this one, and is currently  the only AO system
that can observe this many targets in such a short time~\citep{Terziev2013,Law2014a}.

\subsection{Automated data processing}
\label{sec:lucky}

The data were processed by  automated reduction software  developed for Robo-AO image alignment (see~\cite{Law2014a} and \cite{Law2014b} for the details).  Each  individual frame is corrected  for bias, dark
current and flat field  using standard calibration data, and the frames are then run through the software to co-add them and create a final high resolution image oversampled to $2048\times2048$.  

Stars  in  each  output  image  were  identified  visually  and  their
approximate centers were marked. Relative astrometry and photometry of
wide pairs  where the images  do not overlap  was done by  fitting the
scaled and shifted  image of the brightest component  to the secondary
within 10 pixel radius from the image center. For closer (overlapping)
binaries we used blind deconvolution, as described by \citet{astrom2}.
However,  most  of these  ``blind''  measures  are  superseded by  the
results of speckle processing.

\subsection{Binary-star measurement by speckle processing}
\label{sec:speckle}

Speckle processing is complementary  to the blind deconvolution of images, as it delivers  diffraction-limited resolution even  for low-Strehl data,
when the instantaneous PSF has multiple spikes
(speckles).   The algorithm takes  care of  the multi-speckle  PSF and
works well even at low flux, when the selection of the brightest pixel
for  re-centering lucky  images is  compromised by  the  photon noise.
Moreover, the  power spectrum  (or auto-correlation function,  ACF) is
proportional  to the  {\em square}  of the  signal and,  therefore, it
automatically assigns high weight  to sharp images.  For widely spaced
binaries, speckle  processing is done for  each component individually
to detect close sub-systems.
 
Sub-frames of  256$\times$256 original (un-binned)  pixels centered on
each component  were selected from  the data cubes.  The  background was
estimated  as median  signal  in each  column  of the  full image  and
subtracted, removing the systematic  bias pattern along the CCD lines.
No flat-field  correction was applied.   The average power  spectrum was
calculated on the 256-frame  cube for each selected
component. In parallel,  an image was
produced from these data by re-centering on the selected component and
  weighting individual  frames with  the maximum  intensity  in each
frame. The  weight is therefore proportional to  the image sharpness.
These auxiliary images are not  used for astrometry or photometry, but
are helpful  for verifying companion  detection and for  resolving the
$180^\circ$  ambiguity of  position  angle inherent  to the  classical
speckle processing.

\begin{figure}[!ht]
\plotone{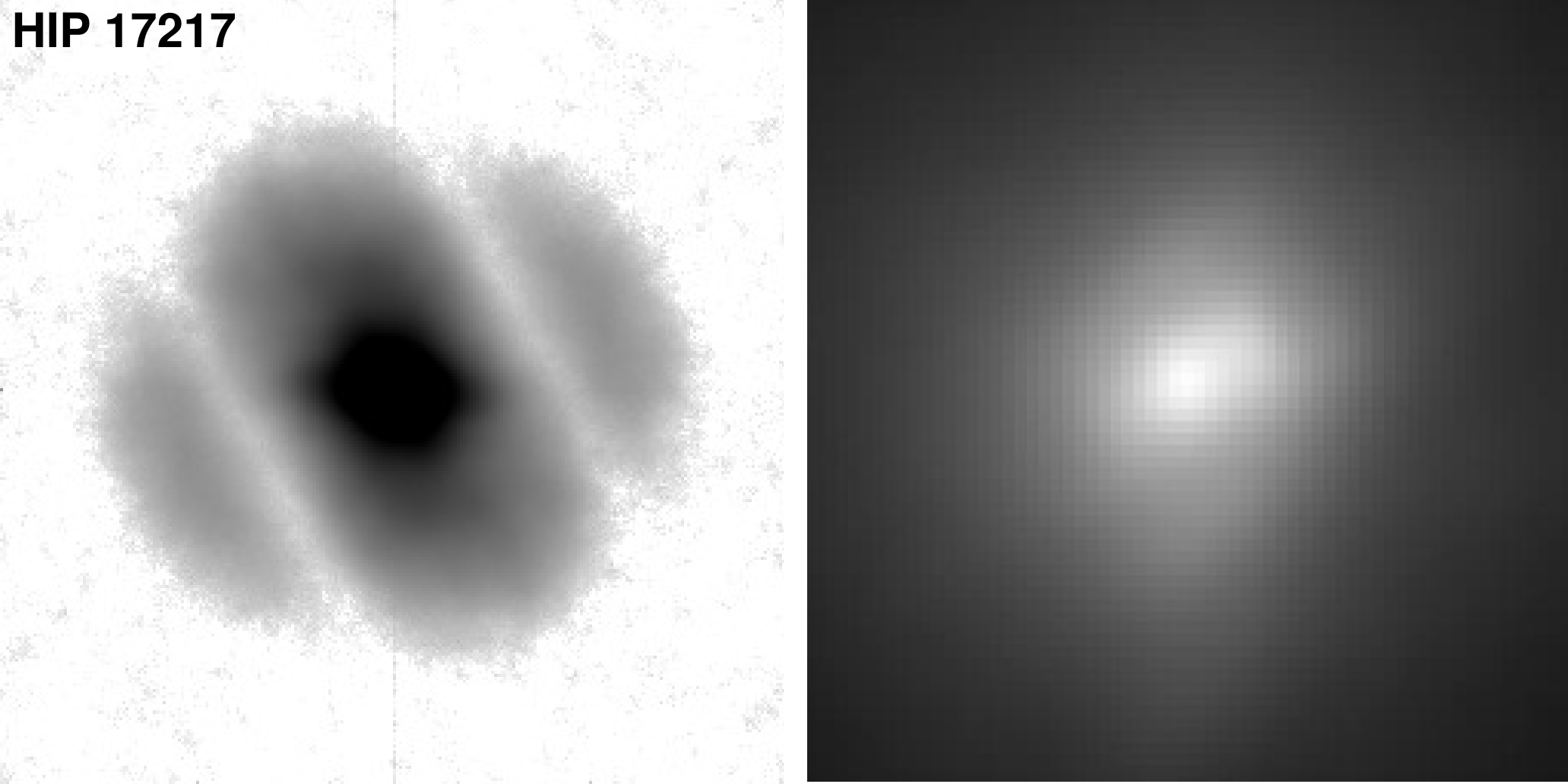}
\caption{Power  spectrum  (left,  inverse  logarithmic scale  from  $10^{-2}$  to
  $10^{-7}$)  and   lucky  image  (right)   of  HIP~17217  (separation
  0\farcs118, $\Delta m = 0.36$).
\label{fig:speckle} }
\end{figure}

The algorithm  of speckle data processing  and binary-star measurement
used      here      is     adapted      from      the     work      of
\citet{TMH10}.  Figure~\ref{fig:speckle}  illustrates  the case  of  a
close   0\farcs12  pair  near   the  diffraction   limit.   Extracting
astrometry and photometry from  the image appears problematic, whereas
the ``fringes''  in the  power spectrum are  very clear  and constrain
 the position and magnitude difference $\Delta m$.

For binaries with nearly equal components, the pipeline image contains
an ``antipode''  because the brightest  pixel on which the  frames are
re-centered may  belong to either  binary component. In this  case the
relative  photometry  derived  from   the  image  is  wrong  (although
correctable), but the speckle  processing delivers the correct $\Delta
m$.  It  also increases the resolution on  distant secondary components
where   the   full-frame      images   suffer   from   the   tilt
anisoplanatism. On the other hand,  if the secondary component is very
faint,  the pipeline technique of re-centering on  the  bright primary  works  much better  than
speckle processing  of the secondary alone.  We  quantify the strength
of the  speckle signal by  its ratio to  the level of photon  noise at
a spatial frequency  two times less  than  the  cutoff frequency,  and
consider  all  measures  resulting   from  the  ``weak''  data  to  be
uncertain.

Speckle processing and imaging are  truly complementary. In the final step,
we   combine   binary-star   measures   from  those   two   processing
techniques. For  binaries with $\rho  < 2''$ the speckle  measures are
preferred (with a few exceptions where the speckle processing failed),
for wider pairs  measures from the images are  retained. Comparison of
repeated measures of the same pairs by speckle and blind deconvolution
shows   their  excellent   agreement  and   gives  an   idea   of  the
precision. Table~\ref{tab:test}  lists average values  and rms scatter
for two sample binaries measured by both methods several times.

\subsection{Detection limits}
\label{sec:det}

\begin{figure}[!ht]
\plotone{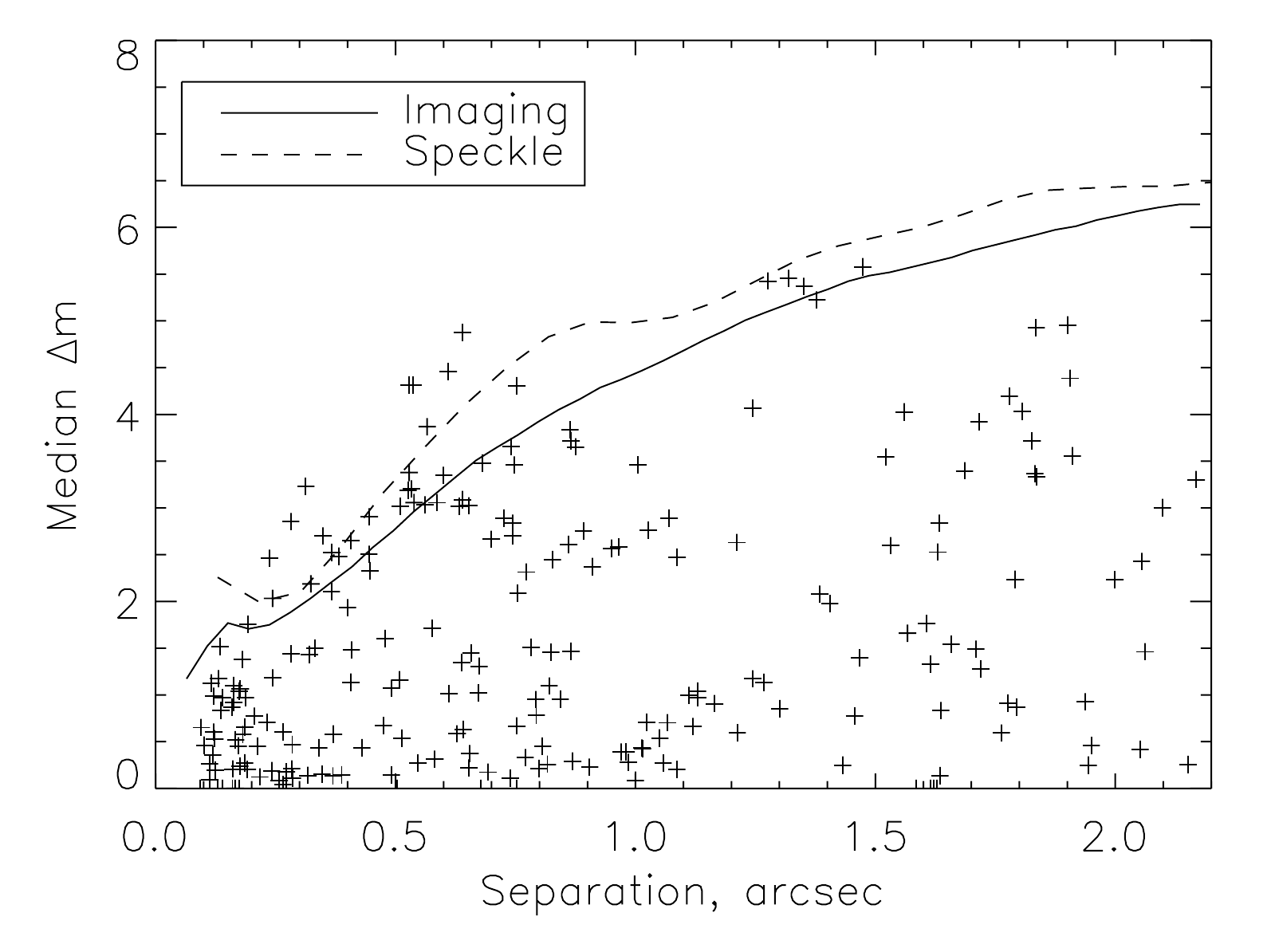}
\caption{Median  detection limit $\Delta  m(\rho)$ determined on the images (full line) and  on the speckle ACFs (dashed line). Measured pairs are plotted as crosses.
\label{fig:det} }
\end{figure}

Limits for  detecting companions in the pipeline  images are evaluated
in  the standard  way.  Fluctuation  of  the signal  in annular  zones
surrounding  the  star  are  computed  and  it  is  assumed  that  the
companions brighter than $5 \sigma$ are detectable. The same method is
applied to  the ACFs in the  speckle processing. The  two estimates of
the detection limits  obtained in this way are  very similar, with the
speckle limits being  normally a little deeper (Figure~\ref{fig:det}).
Depending on  variable AO  correction and target  brightness, the
individual  detection limits  vary  substantially.  For each target, we  list the  best
(deepest) limits at  three characteristic separations  of 0\farcs15, 0\farcs8
and  2\farcs1,  selecting  the  best  of  two  alternative  processing
techniques.  When  the target was observed several  times, the deepest
detection limits are reported in the table of final results.

All of the survey primary targets  are bright, but some secondary  components present a
problem as they are  too  faint to  provide a useful signal  in the  power
spectrum. Their  pipeline images  can show a bright  1-pixel spike  at the
center  surrounded  by  a  fuzzy   halo; this   is  created  by
re-centering  on  the photon  events,  rather  than  on the  brightest
speckle,  so  the  resolution  is  lost.   Detection  limits  computed
formally from such  images are over-estimated. Yet, we  prefer to keep
such  weak data  in  the  final table,  despite  this obvious  caveat,
because they still contain some  information on the duplicity.

\subsection{Calibration and distortion correction}
\label{sec:dist}

The re-imaging system of the Robo-AO instrument contains a double optical
relay  with  off-axis parabolic  mirrors.   Such  relays  are known  to
introduce  quadratic image  distortion.  In  the case  of  Robo-AO, the
distortion, as reported by the  optical design, is quite strong. It is
directed along the visible CCD  columns and displaces all sources  down by as much
as 26 pixels in the corners.

Independently  of the  optical design,  the distortion  in  Robo-AO was
mapped  on  the sky  using  the image  of  the  globular cluster  M~15
(S.~R.~Hildebrandt,  private communication, 2012).  The  pixel scale  in the
X-direction was found to be 43.74~mas, while the scale in Y
was slightly  different. The orientation is  such that the  +Y axis of
the detector points  at a position angle 23\fdg9  from  North
and  the   +X  axis  points   East  (the   image  has  mirror
orientation). The quadratic  distortion agrees with the optical-design
data and is directed along Y.

If we express the actual $X,Y$  star coordinates in the  image in the
``small''  over-sampled  pixels of the automatic pipeline reduced image, the  undistorted  coordinate $Y'$  is
determined as
\begin{equation}
Y'   = Y - 0.0299   \Delta  Y  + C_X (\Delta X)^2 +  C_Y  (\Delta  Y)^2   ,
\label{eq:dist}
\end{equation}
where $\Delta X$  and $\Delta Y$ are counted from  the pipeline image center at
(1024,1024) and the  coefficients are $C_Y = -  2.54\;10^{-5}$, $C_X =
-2.55\;10^{-5}$.  The  linear term in $Y$ corresponds  to the modified
vertical scale. It  is caused by the mismatch  between the optical and
CCD centers combined with the quadratic distortion.

This distortion  substantially affects  the measurements of  even close
binaries because  the CCD  lines projected on  the sky  are parabolic.
The distortion-induced  tilt of the  lines reaches 5\fdg6 in the
corners of  the CCD.   We correct the  geometric distortion  using the
pixel    coordinates     of    each    source     in    the    images.
Equation~\ref{eq:dist} is applied to  each component of a binary.  The
parameters  $\rho, \theta$  are re-calculated  from the  difference of
un-distorted pixel  coordinates $X,Y'$.  In some  instances the angles
of close binaries are changed by as much as $2^\circ$ after distortion
correction.  The  scatter of  relative positions of  binaries measured
several  times  is  reduced  after distortion  correction,  while  the
relative  positions of  wide pairs  became closer  to  their positions
derived from 2MASS.

\subsection{Caveats}
\label{sec:prob}

As with any observing program, there were issues with some of the data collected that required some extra effort to make it useful.  Each pipeline image was examined manually to find and remove erroneous measurements or detections; a graphical IDL tool was created to do this, and also used to mark the components and fit wide binaries.  Owing to the P60 pointing errors, some images were  empty, or the targets
were found close  to the edge of the FOV (see  \S~3.1). 

The quality of
AO compensation, Strehl ratio, and the  width of the  PSF depended on
the seeing conditions and  varied substantially. As a consequence, the
depth of companion  detection was also variable. In  some instances we
detected new companions in the good-quality images, but missed them in
the poor ones. The PSF had a persistent structure (``static speckle'')
which could be mistaken for  a companion. This structure also depended
on   the   AO   compensation   quality   and,   possibly,   on   other
factors. Fortunately,  there were always  other stars observed  in the
same conditions so the PSF could be compared in order to verify suspicious detections.

In a few  cases, the pipeline algorithm occasionally  centered a frame
on a  bright spike caused by  cosmic rays, while  the remaining frames
were centered correctly. This produced a false satellite that could be
mistaken for a new companion. Also, several faint satellites caused by
internal  reflections in the  optics are  always visible  around bright ($V \lesssim$ 3 mag)
targets.   These  reflections effectively  reduce  the  dynamic range  of
companion detection.  We  were able to reduce some images  where the  primary component  was saturated.
Such   data    are   still    useful   for   detecting    wide   faint
companions. Relative  astrometry and  photometry of such  binaries was
done by  fitting PSF  within an annulus,  i.e.  excluding  the central
saturated PSF core from the fit.  

The accuracy of  the measured positions of wide  binaries is less than
for  the close  ones.   This is  likely  caused by  the residual  tilt
anisoplanatism: the differential tilt  between the components caused by
high-altitude turbulence  is not completely averaged  during the short
60~s exposure.

In many cases, we were able to re-observe targets after the data analysis showed that the data were not sufficient; the ability of Robo-AO to observe quickly and efficiently allowed many objects to be added to the final analysis that would have not been available otherwise.

\section{Observing data tables}
\label{sec:data}

The list  of all components  successfully observed in this  program is
given  in  Table~\ref{tab:exprobo_short}\footnote{Full  versions  of  Tables~\ref{tab:exprobo_short}  and~\ref{tab:double_short}  are
  published on-line}.   The list contains 695  components belonging to
595 systems.  Its  columns contain the HIP1 designation (which is the {\em Hipparcos}  number of the
primary component  in each system, and is used as the main identifier through this paper),  component designation, equatorial
coordinates  of the  observed component,  its $V$  magnitude,  and its
separation  from the  primary  (these  data are  taken  from the  main
database  of the  FG-67 sample).   Secondary components  have non-zero
separations.  The  following columns list the date  of observation and
detection limits at separations  of 0\farcs15, 0\farcs8, and 2\farcs1.
The last column gives the distance  from the star to the nearest frame
border,   indicating  the   maximum  separation   of   its  detectable
satellites.

Measurements of resolved doubles are  given in Table~~\ref{tab:double_short}\textsuperscript{1}.  Each pair is
identified by its  WDS code \citep{Mason2001}, discoverer designation,
and component  designation. In column 3, the HIP1 number of the
  primary  star in  each  system  is given.   Following that are the  date  and number of  observations
(multiple  observations  made in  the  same  filter   within  0.2~yr  are
averaged), position angle (degrees), position angle error (degrees), separation (arcseconds), separation
error(milliarcseconds),  and magnitude difference.   The errors  of position  angle and
separation are computed as errors of  the mean in the case of averaged
data, otherwise estimated from  the speckle processing, or listed as zero
for PSF fitting of wide pairs. Uncertain measures (weak speckle signal
or blind  deconvolution) are  marked by colons  after $\Delta  m$.  We
also  mark  the  bandpass when  it differs  from $i'$  (the LP600
filter is  denoted by $w$).  The  last three columns  give deviations in
position angle and separation from the orbits, when available.

\section{Results}
\label{sec:res}

\subsection{Newly resolved systems}
\label{sec:new}

\begin{figure*}[p]
\center
    \includegraphics[scale=0.5,angle=90,natwidth=704,natheight=267]{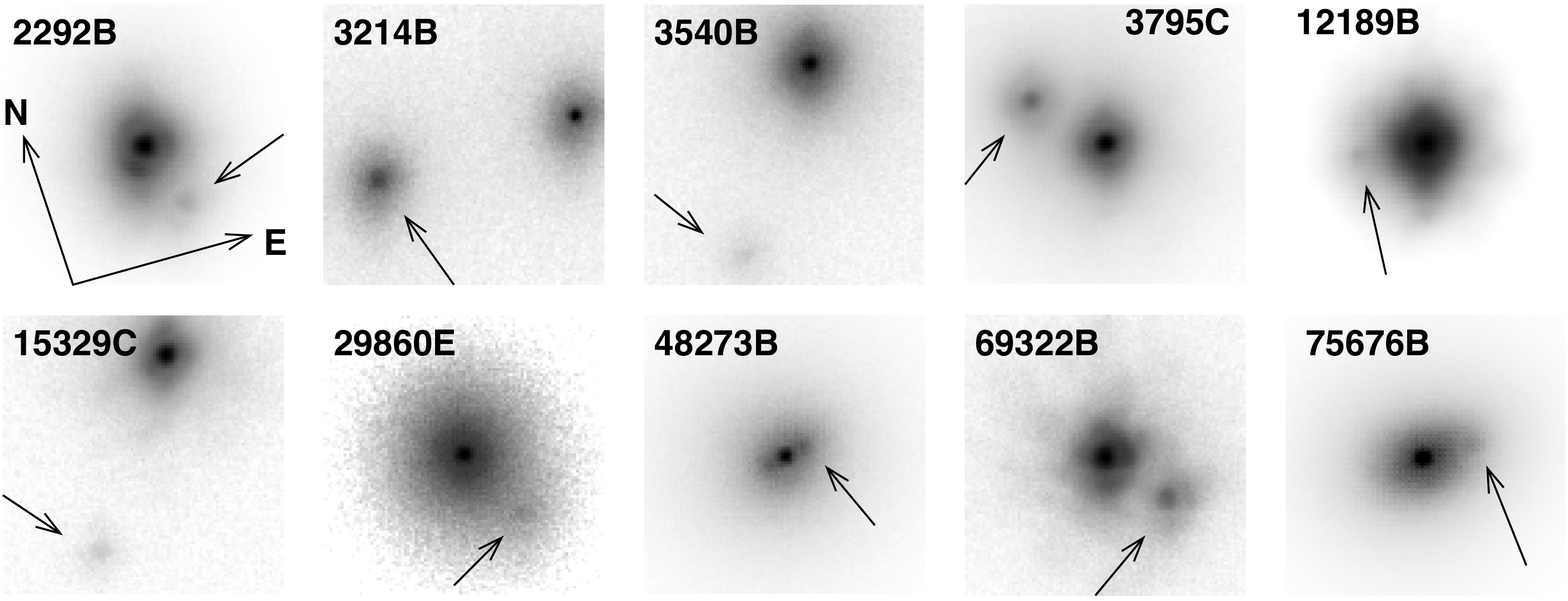} 
    \includegraphics[scale=0.5,angle=90,natwidth=702,natheight=266]{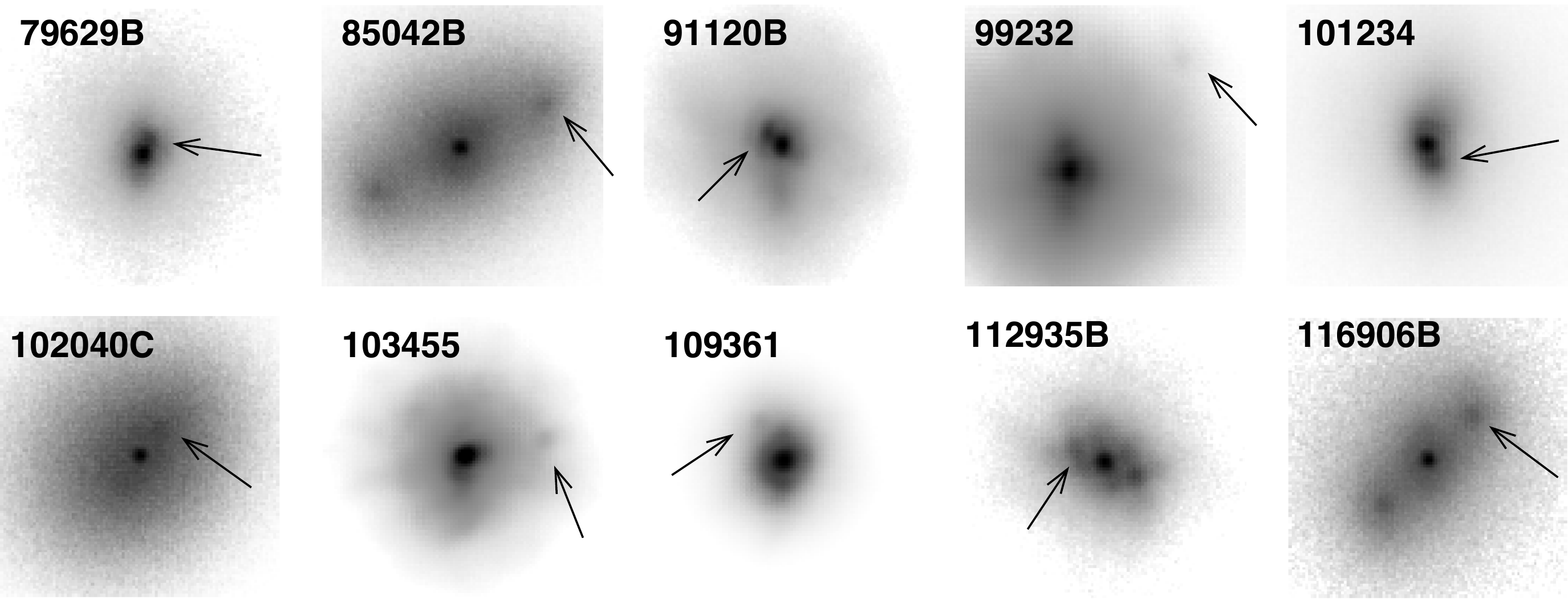}
\caption{Images   of   some  new    binaries  discovered   with
  Robo-AO. Each fragment of the pipeline image has a size of 100 pixels
  (2\farcs2),  the  intensity  scale  is  adjusted  to  highlight  the
  companions, indicated  by arrows. The images  are displayed without
  rotation and labeled by HIP numbers. 
\label{fig:img} }
\end{figure*}

\begin{figure*}[p]
\plotone{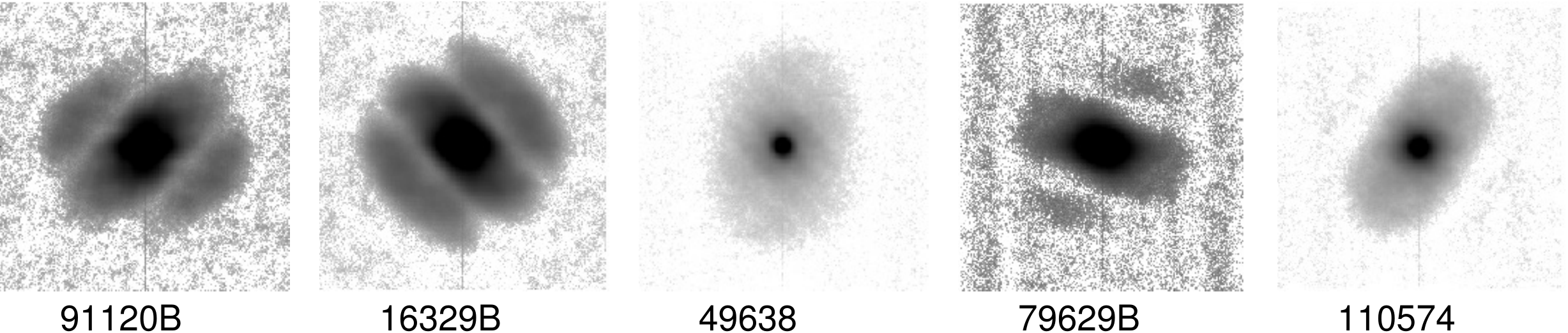}
\caption{Power spectra of some newly resolved close binaries. Primaries are identified by their HIP1 number.
\label{fig:power} }
\end{figure*}

The  list of  newly  resolved binary  companions (including  unrelated
background  stars) is  extracted  from the  main  table and  presented
separately in  Table~\ref{tab:new}; they are  given the ``discoverer code''  RAO in
the WDS; a space is added between discoverer codes and component designations to make it clear they are not the same (i.e. component designations such as Aa and discoverer codes such as RAO 20 are parallel and must not be confused).   Some binaries were  measured several times,  confirming the
measured resolution.    Images   of   new   close  pairs   are   presented   in
Figure~\ref{fig:img}.  The  closest pairs are detected  in the speckle
processing by examining the power spectra; these cases are illustrated
by Figure~\ref{fig:power}.  The new wide physical pairs are confirmed
by  comparing the measured  positions of  the  components with  their
position in the  2MASS catalog, and by considering additional Êinformation Êsuch as Êmagnitude, separation, color, crowdedness Êof the field, and proper Êmotion of the main Êtarget. ÊÊWe Êdo Ênot Êcompute the formal Êprobability Êof Êphysical association based Êon relative astrometry because Êthe Êmotion can be Êdistorted by subsystems, Êwhile chance projections Êwith small and similar motions do happen sometimes (see e.g.~\citet{LEP}). Other Êfactors involved in the classification Êare difficult to quantify in Êterms of probability. ÊÊMost faint companions Êin Table~\ref{tab:new}~wider that Ê10\arcsec~are Êoptical with a Êlarge confidence; only three of them have an uncertain status.  In some instances,  the companions are
not  found in  the catalog,  but the  2MASS images  in the  $K_s$ band
reveal  their presence  as  blends.  This  is  taken as  confirmation,
because the  companion position  matches (at least  qualitatively) and
because  the  detection of  the  blend  in  2MASS indicates  that  the
companion is brighter in $K_s$ than in $i'$, indicating it is a cooler object below the detection threshold in the catalog.  

Notes in the Appendix give additional information on
the multiple systems appearing in  Table~\ref{tab:new}, the status of the measured
binary pairs ('P'  for true physical binaries, 'O'  for chance optical
alignments   of   unrelated   stars),   and  the   reasons   for   this
classification. Follow-up observations  in September 2013 with the
PALM~3000 AO system at the  5-m Hale telescope at Palomar Observatory~\citep{Dekany2013} are mentioned
where  relevant, while  their  full  results will  be  published in  a
forthcoming paper (Roberts et al, in preparation).  Two of the systems with newly resolved binaries, HIP 2292B (= HIP 2350) and HIP 12189B (= HIP 12184), are also exoplanet hosts; a complete discussion of their properties are included in~\cite{Roberts2014b}.    Two RAO pairs
were  also  measured  by  speckle  interferometry at  the  SOAR  4.1-m
telescope \citep{Tokovinin2014b}.

\subsection{Binarity of secondary components}
\label{sec:sec}

\begin{figure}[!ht]
\plotone{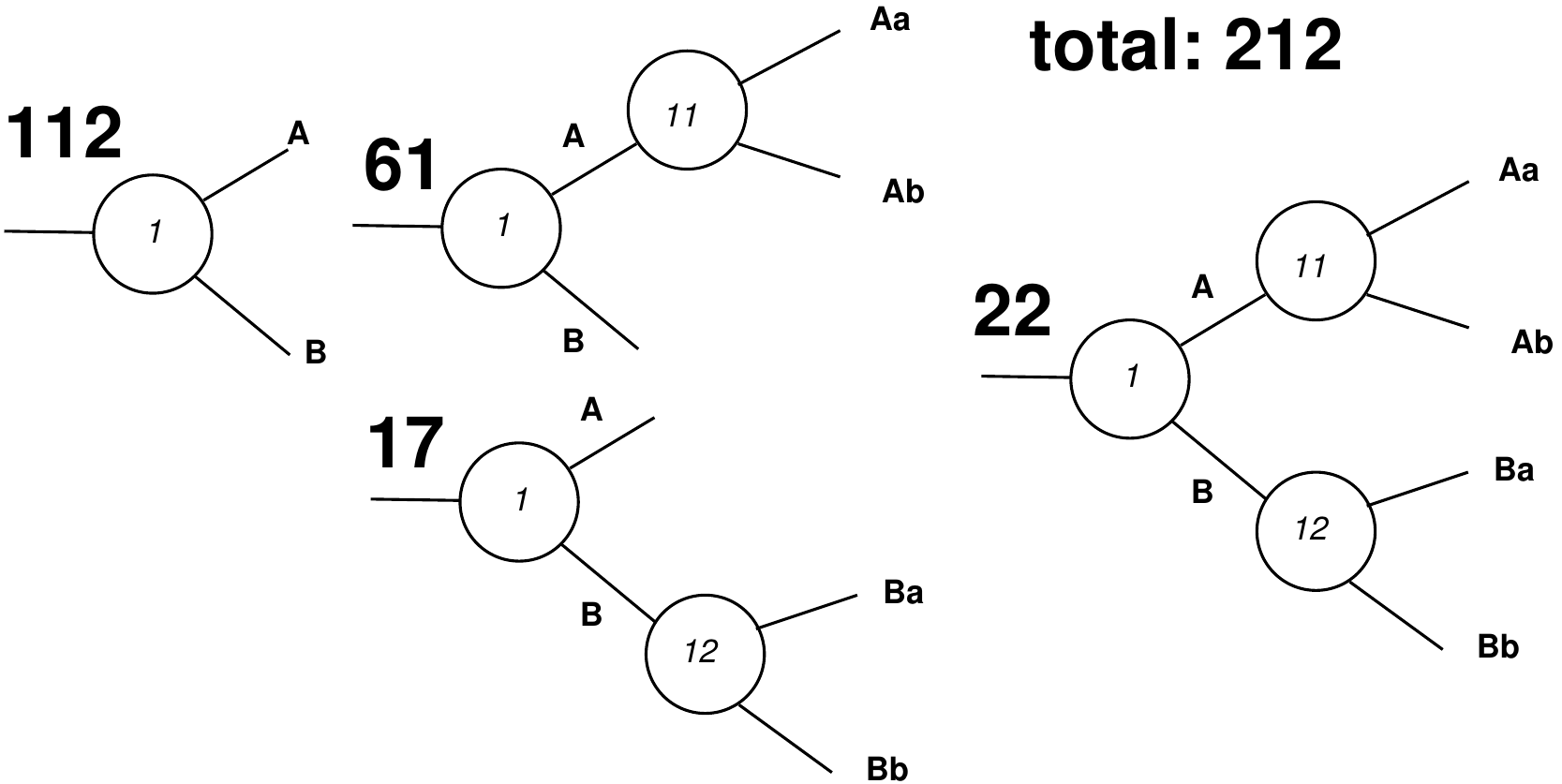}
\caption{Observed  numbers of different  hierarchies in  212 wide binaries
  surveyed by Robo-AO.
\label{fig:sys-count} }
\end{figure}

Overall,  there are  212 secondary  components with  separations above
$10''$  that were  observed with  Robo-AO.  Figure~\ref{fig:sys-count}
illustrates different  hierarchies found in this  sub-sample. The wide
binary is the root of the  hierarchy (level 1); in 112 cases there are
no sub-systems.  Another 100 wide binaries contain inner hierarchies of
level 11  (sub-system in the primary component),  level 12 (sub-system
in the secondary) or both (that is, a 2+2 quadruple). Here we silently
ignore  several sub-sub-systems  (levels 111  etc.),  and  restrict the
discussion  to  the  hierarchical  levels  1, 11,  and  12.   As  some
subsystems are most likely not yet discovered,  it is safe  to say that  at least
half  of  those wide  binaries  are  in  fact triple  or  higher-order
multiples.

Orbital  periods of  wide binaries  are estimated  roughly  from their
projected  separation and denoted  as $P^*$  to distinguish  them from
orbital solutions.  Masses of binary components and the mass ratio ($q=
{\cal M}_2/{\cal M}_1 $)  are estimated from their absolute magnitudes,
assuming  that the stars  obey standard  relations for  the main sequence.
The  detection  limits of  imaging  are  converted  from the  observed
parameters $(\rho,  \Delta m)$ into binary parameters  $(P^*, q)$, and
combined  with  the  detection  limits  from  spectroscopy  and  other
techniques  \citep[more details in][]{Tokovinin2014a}.

\begin{figure}[!ht]
\epsscale{0.75}
\plotone{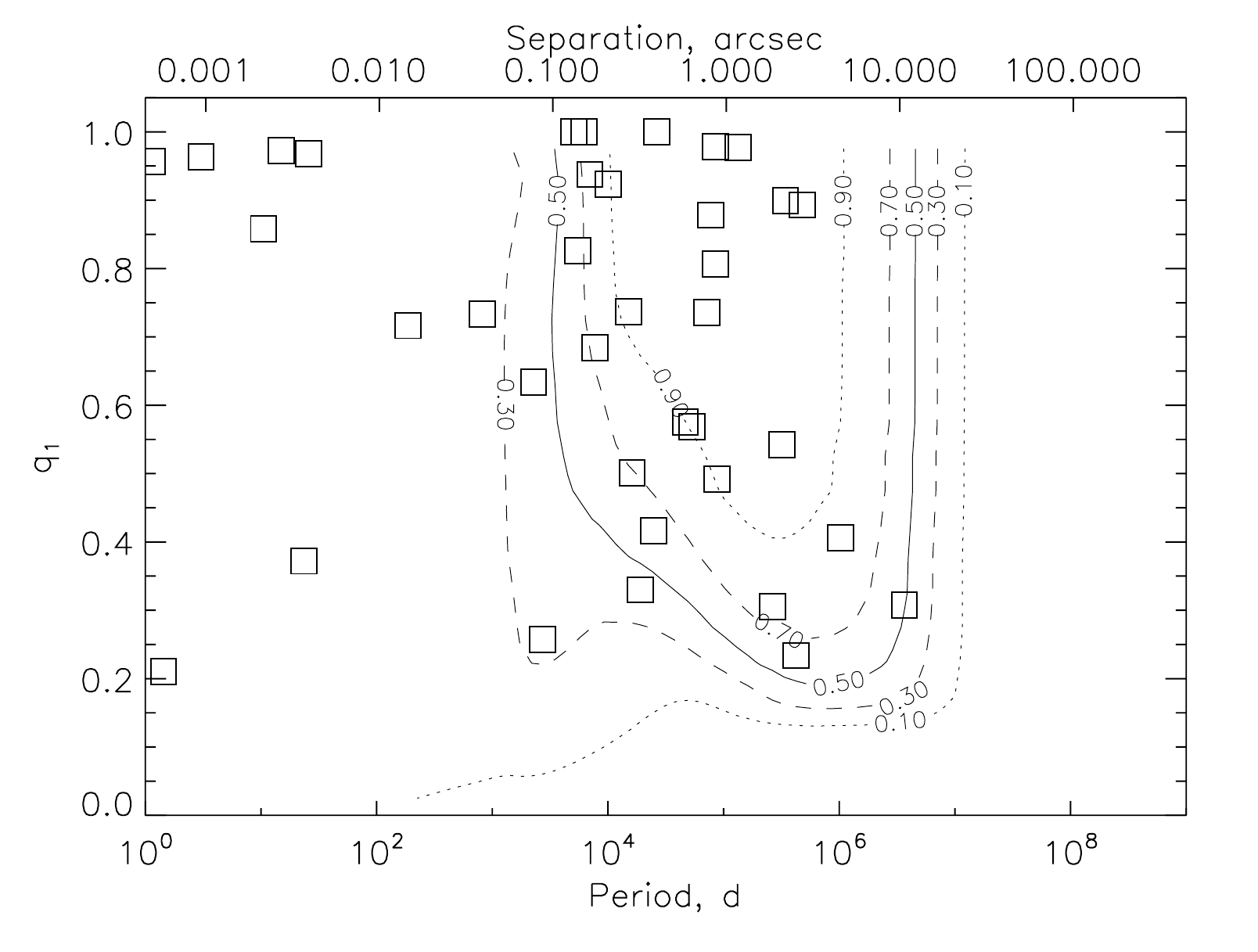} \plotone{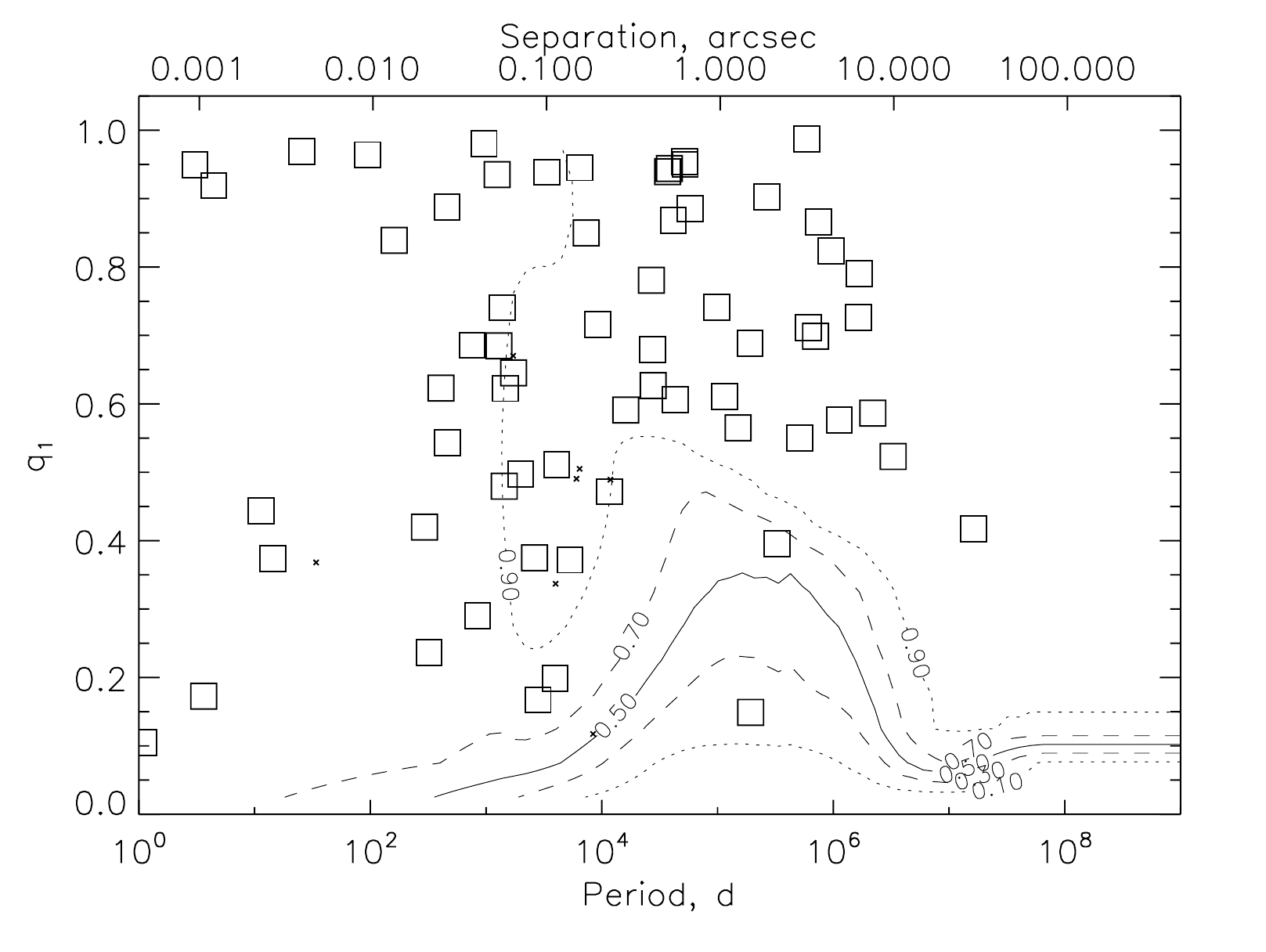}
\caption{ Distribution  of 39 secondary sub-systems  (upper panel) and 83 primary sub-systems (lower panel) of wide binaries in the $(P,q)$ plane.   The contours indicate  average detection  probability.  The angular separation on the upper  axis corresponds to the distance of 50~pc.
\label{fig:pqsec} }
\end{figure}

There  are  83  primary  (level   11)  and  39  secondary  (level  12)
sub-systems in the surveyed objects (Figure~\ref{fig:pqsec}); nineteen new secondary subsystems were found, doubling their previously observed fraction.  At first
glance, the primary sub-systems still dominate, but remember that most of the 
primary  targets (bright  F-  and G-dwarfs)  were  surveyed in  radial
velocity  (RV), while  only a  few brighter  secondaries have  such RV
coverage.   The  primaries  also  benefit  from  the  {\em  Hipparcos}
discoveries of  acceleration sub-systems \citep{MK05}, which explains
deeper  detections at  $P  \sim  10^3$~d.  At  $P  \sim 10^6$~d  the
situation is reversed in favor  of secondaries, owing to their imagery
with  Robo-AO.  At  still longer  periods (separation  $\ga  5''$), the
companion  census   for  primaries   is  complete  because   of  2MASS
\citep{Tokovinin2011} and CPM \citep{LEP} data, while for the secondaries
the detections are  restricted by the size of the Robo-AO field.  However,
this  makes  little  difference   because  extremely  wide  sub-systems  are
dynamically unstable.

 It is convenient to express orbital
periods by x, the logarithm of the period in days.  The criterion
of dynamical stability by \citet{Mardling2001} requires that the ratio of
outer  to  inner  periods  in  triple systems  be  greater  than  4.7,  
depending   on  the eccentricity  in  the outer  orbit.  The  {\em
  dynamical  truncation} function  $F(\Delta  x)$ (where  $\Delta x  =
x_{\rm out}  - x_{\rm in} $ is  the logarithm of the  period ratio) is
the probability of a given  triple system being dynamically stable; it
is modeled  here as zero  for $\Delta  x < 0.7$,  one for $\Delta  x >
1.7$, and linear in-between.   Considering the large range in periods,
the exact form of this function has little influence on the results.

In  the  sub-sample  of  212  wide  binaries,  the  average  dynamical
truncation is 0.95 at inner period $x_{\rm in} = 5$ and 0.5 at $x_{\rm
  in} = 6$,  explaining the lack of wide sub-systems  at levels 11 and
12. There is  no need to look for  binary secondaries with separations
$>10''$ because those are  intrinsically rare.  The limitation imposed
by the Robo-AO field of view is therefore not important for this work.

We select  for analysis the  period range $3.5  \le x_{\rm in}  \le 5$
where the dynamical  truncation is not important and  the detection of
secondary sub-systems  with Robo-AO is relatively  complete. This range
corresponds to  1 decade  in separation. We  compare the  frequency of
primary  and secondary  sub-systems  in this  range  of periods  among
selected  212 wide  binaries  in Table~\ref{tab:sec}.   The number  of
missed  companions obviously depends  on the  distribution of  the mass
ratio, modeled as a power  law $f(q) \propto q^\beta$. It is generally
accepted that the  power index $\beta$ for   solar-type binaries is
close   to   zero   (uniform   distribution)  or   slightly   negative
(Duch\^ene \& Kraus 2013), while low-mass binaries tend to have more equal
components and $\beta \sim 1$.

The probability of detecting a sub-system averaged over period (in the
selected interval) and over the mass  ratio $q$ can be used to correct
the  raw  companion  frequency,  if  $\beta =0$.   However,  the  data
collected here indicate that  in the secondary sub-systems $\beta \sim
1$, requiring a smaller correction for missed binaries. Therefore, the
estimated frequency of secondary  sub-systems with $3.5 \le x_{\rm in}
\le 5$ is $0.13\pm 0.03$, assuming conservatively $\beta = 1$.  It is only slightly less than the frequency
of  primary (level-11)  sub-systems, which  is less  sensitive  to the
assumed $\beta$ and is somewhere between 0.13 and 0.15
(Table~\ref{tab:sec}). 

The  frequency  of sub-systems  found  here  is  close to  the  
frequency  of solar-type  binaries in  the same  period range.   For the
log-normal period distribution and the binary fraction of 0.46 derived
by  \citet{Raghavan2010}, the  fraction  of companions  with $3.5  \le
x_{\rm in}  \le 5$ is  0.112 (a similar  estimate is obtained  for the
full FG-67 sample).  The frequency of sub-systems in the components of
wide binaries may  be enhanced in comparison to  the single stars, but
only slightly.

The actual frequency of sub-systems  in secondaries may be even higher
because  of  two  additional  factors.   First, the  methods  used  to
identify  wide  secondaries  introduce  some selection  against  close
binaries \citep{LEP}. Second, detection limits here may be
over-estimated   (see \S\ref{sec:det}).    When  the   assumed
detection limits are deeper than in reality, the correction of the raw
frequency for incomplete detection becomes smaller.

Quite surprisingly, the sub-systems in primary and secondary are often
found at the same time.   Among the 39 secondary sub-systems, 22 (more
than  half!) belong  to  2+2 quadruples  (Figure~\ref{fig:sys-count}).
Considering  that many  secondary  sub-systems with  short periods  are
still missed,  the actual  fraction of 2+2  quadruples among  212 wide
binaries  should be  more  than 10\%.   This surprising correlation likely has a root in stellar system formation processes and requires further study.

\subsection{Frequency of tertiary components}
\label{sec:tert}

In the second  part of our survey we looked  for distant companions to
close  binaries.   In other words, the problem  is  ``reversed''  and  we explore  the
hierarchy from inside-out. Similar studies were made by \citet{Tok06},
\citet{Allen2007}, \citet{Rucinski2007}, and others.

The second survey  selects main targets which are  close binaries with
$P < 100$~yr ($x < 4.56$). Binaries with yet unknown periods, such as
acceleration  binaries   from  {\em  Hipparcos}   \citep{MK05}  and/or
spectroscopic  binaries from  \citet{Nordstrom2004}, are  included. We
  resolved  three  acceleration  binaries (HIP  6653,  101234,
103455) with Robo-AO,  while more were resolved by a  targeted campaign with
the   NICI  instrument   \citep{astrom1,astrom2}.  However,   as  both
acceleration and  RV techniques  involve a non-negligible  false alarm
probability, some of those presumed  close binaries are in fact single
stars.

For the purpose of statistical analysis, we restrict the sample to 241
inner binaries  with {\em known}  periods $P <100$~yr.  In  the cases
where even closer inner sub-systems are present (e.g.  the A-component
of a visual binary is  a spectroscopic pair), we select the inner-most
(closest)  pair and  proceed  outwards.  The  71  close binaries  have
additional outer (tertiary) companions; some
of the tertiary components are  themselves close binaries (in 2+2 quadruples),
and  sometimes  there  are  even more  distant  companions  (3+1
quadruples).   The  raw  frequency  of  tertiaries is  29\%.  When  we
restrict the sample to shorter  inner periods, the fraction of triples
increases, but  the sample becomes smaller and  the statistical errors
increase.

\begin{figure}
\plotone{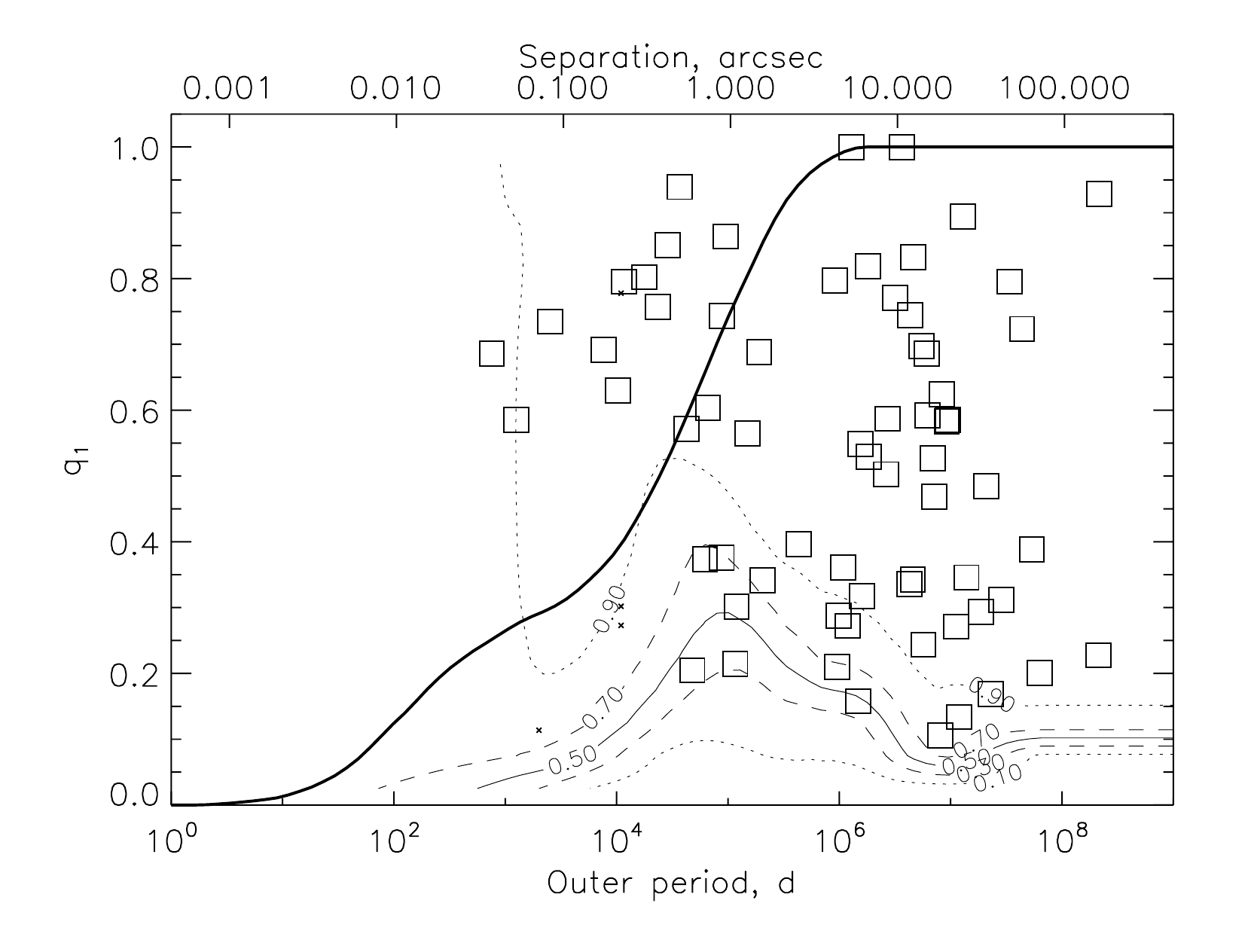}
\caption{ Periods  and mass  ratios of 71  tertiary components  to 241
  binaries with  $P^* < 100$~yr observed with  Robo-AO.  The contours
  indicate average detection probability. The full line over-plots the
  average dynamical truncation (not to be confused with $q$).
\label{fig:pqout} }
\end{figure}

Figure~\ref{fig:pqout}  presents tertiary  components in  the $(P_{\rm
  out}, q_1)$ plane, where $q_1  = {\cal M}_3/{\cal M}_1$ is the ratio
of  the   masses  between   tertiary  and  primary   components.   The
iso-detection curves  go deeper than  for the general  sample (compare
with  Figure~\ref{fig:pqsec},  lower   panel)  owing  to  the  Robo-AO
imaging.  Many of tertiary components  are found in the 2MASS catalog,
but their colors alone were  not sufficient to establish that they are
physical, especially in crowded fields.  The second-epoch imaging with
Robo-AO  helps  here.  A few  uncertain  cases  (companions detected  by
Robo-AO in crowded fields without 2nd epoch data) are not yet accepted
as real.

The   average   dynamical  truncation   curve   $F(x_{\rm  out})$   is
over-plotted.   It explains  why   tertiaries with  short periods
$x_{\rm  out}  \le  3$  are  rare. Such  tertiaries  would  be  easily
detectable  by spectroscopy  and  their scarcity  is  genuine.  As  in
\S\ref{sec:sec},  we evaluate the  frequency of  outer systems  in the
outer period  range of  1.5~dex, $6  \le x_{\rm out}  \le 7.5$,  -- the
range free from dynamical truncation and well covered by the detection
techniques.  It  is safe to  assume $\beta=0$ for wide  components, in
which  case  the  average  detection  probability is  0.88.   The  raw
companion  frequency of  $34/241  =0.141 \pm  0.024$  is corrected  to
$0.161 \pm 0.028$.  Owing to the deep detection limit, there is little
difference  between the  raw and  detection-corrected  estimates.  The
frequency is significantly higher than the companion frequency to main
targets  in the  same period  range, 0.090  
\citep[again using  the  data of][]{Raghavan2010}.

It is  premature to speculate about  the meaning of  this finding.  In
Figure~\ref{fig:pqout} we  note   the   ``pile-up''  of   tertiary
companions at $x_{\rm out} \sim 7$ (separations on the order of $10''$ and
projected separations on the order of 500~AU). If this feature is not
a statistical  fluctuation, it might imply  some characteristic scale
in the formation of solar-type  multiple stars. 

The overall raw frequency of tertiary components (with all periods) is
only 29\%.   If we assume  that the intrinsic distribution  of $x_{\rm
  out}$  is Gaussian  \citep{Raghavan2010} and  that their  mass ratio
$q_1$  is  distributed  uniformly,  the companion  frequency  of  this
underlying  distribution  (before  applying dynamical  truncation  and
detection filters)  is determined  to be 0.54;  i.e., slightly  but not
dramatically  enhanced  in comparison  with  0.46  for all  solar-type
dwarfs.  Most binaries with $P < 100$~yr  do not  have outer tertiary  companions; a similar  conclusion was
reached by  \citet{Tok06} for  spectroscopic binaries with  periods of
$\sim$10~d and longer.  On the  other hand, the closest binaries, with
$P <3$~d, are found almost exclusively in triples.

\subsection{New and updated orbits}
\label{sec:orbits}

For  purposes  of  orbit  determination,  Robo-AO  provides  a  unique
collection   of   data.    While   there  are   certainly   exceptions
\citep{Tokovinin2012,Horch2012}, the majority of interferometric data,
whether with a filled or dilute  aperture, have modest differential magnitude
limits of  about $\Delta  m = 3$ mag.  For  many pairs of  larger magnitude
difference, the Robo-AO observation represent the first observation by
a  high-resolution  technique,  and  given the  decline  in  classical
micrometry, the first  observation of any kind in  many decades. Given
these  factors, the  number  of first  orbits  as well  as those  with
significant   changes  from   the  previous   determinations   is  not
surprising.

\begin{figure}[p]
\begin{center}
\epsscale{1.00}
\plottwo{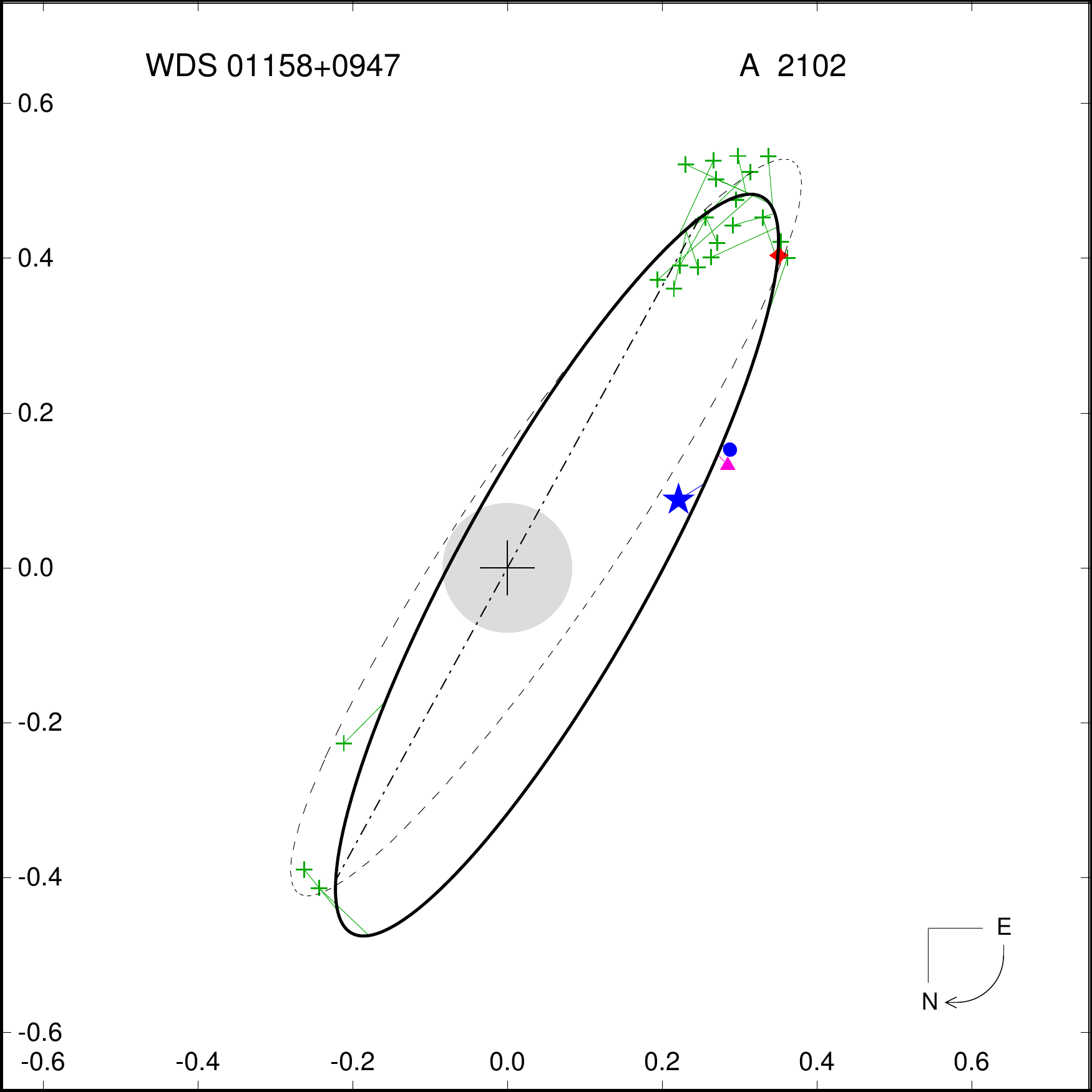} {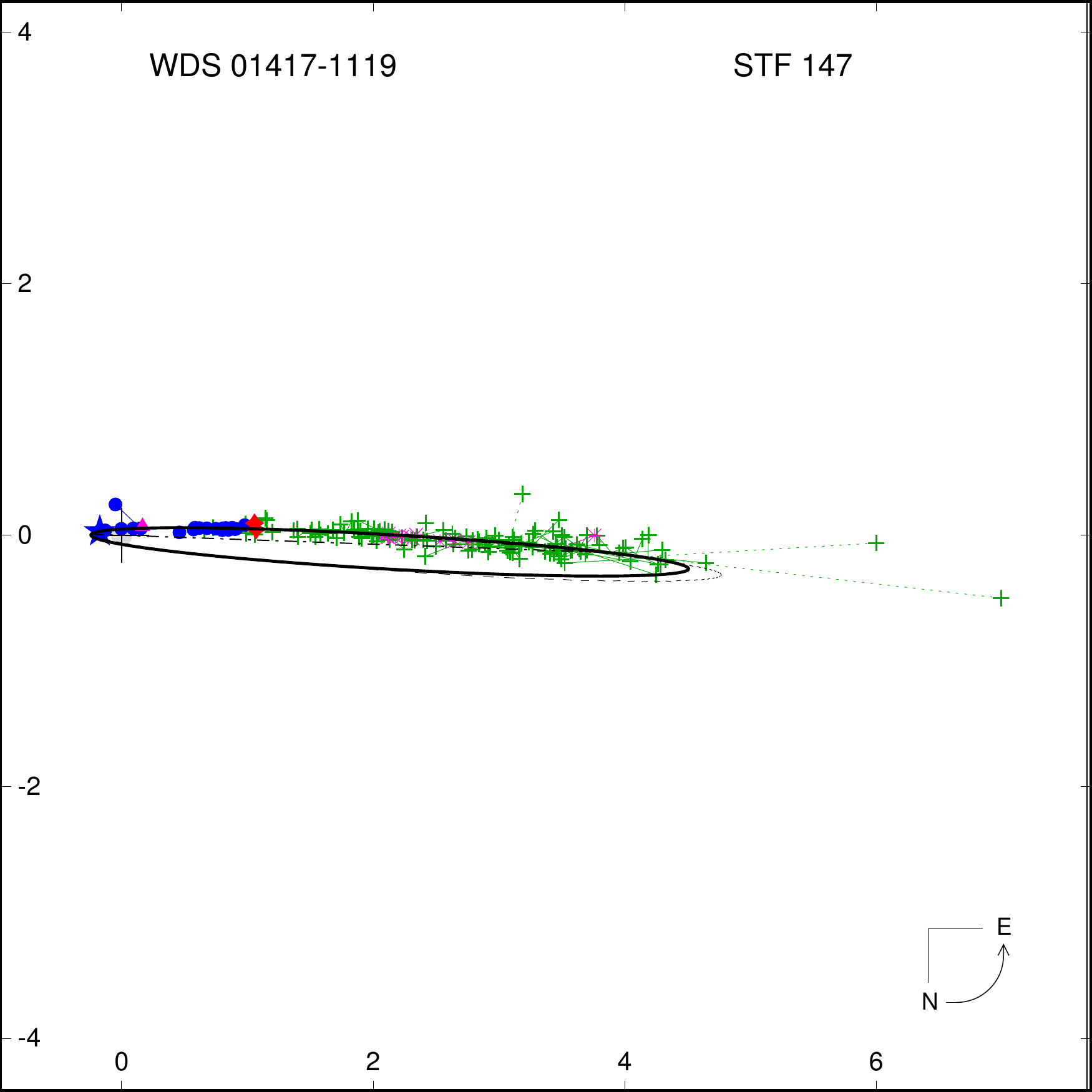}
\vskip 0.05in
\plottwo{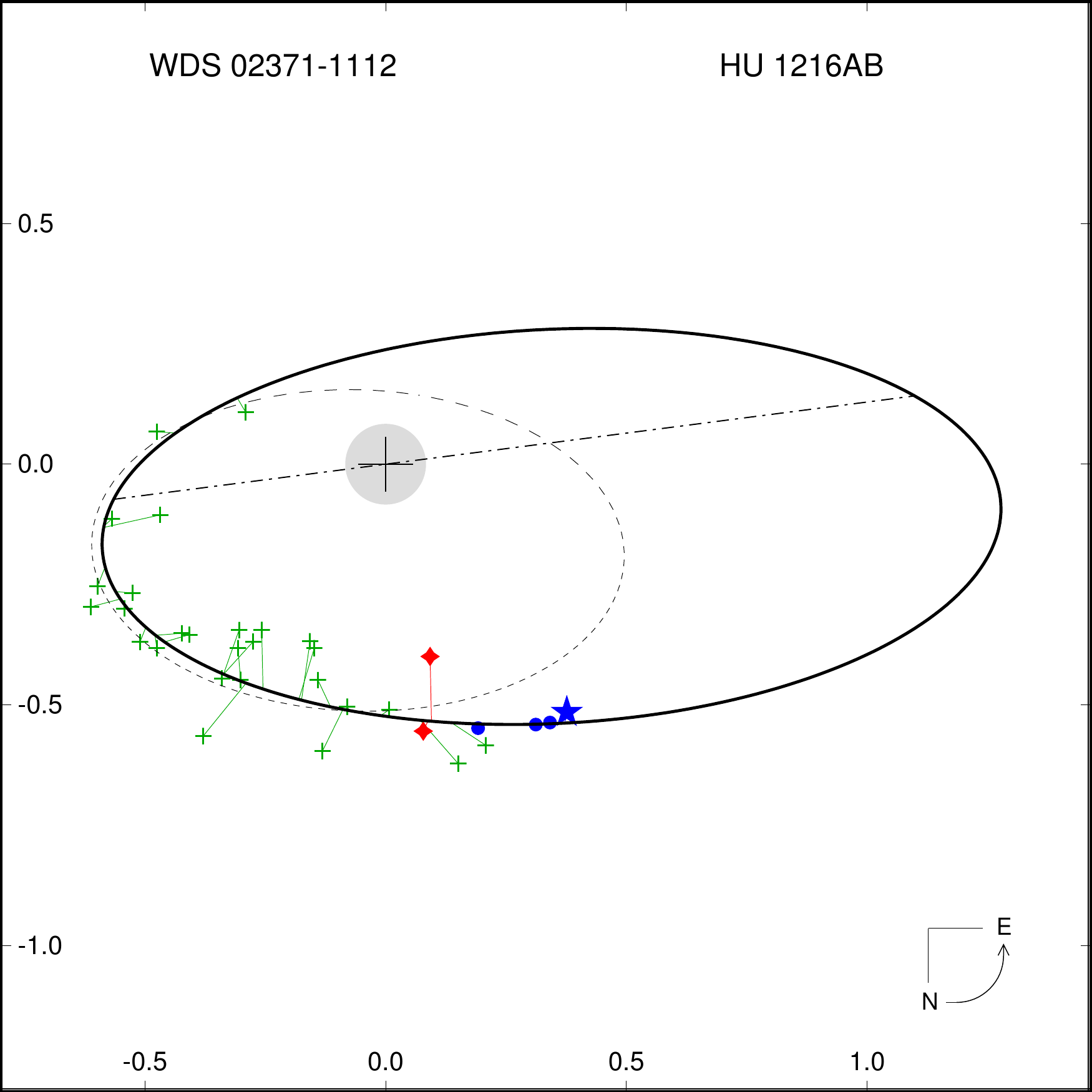}{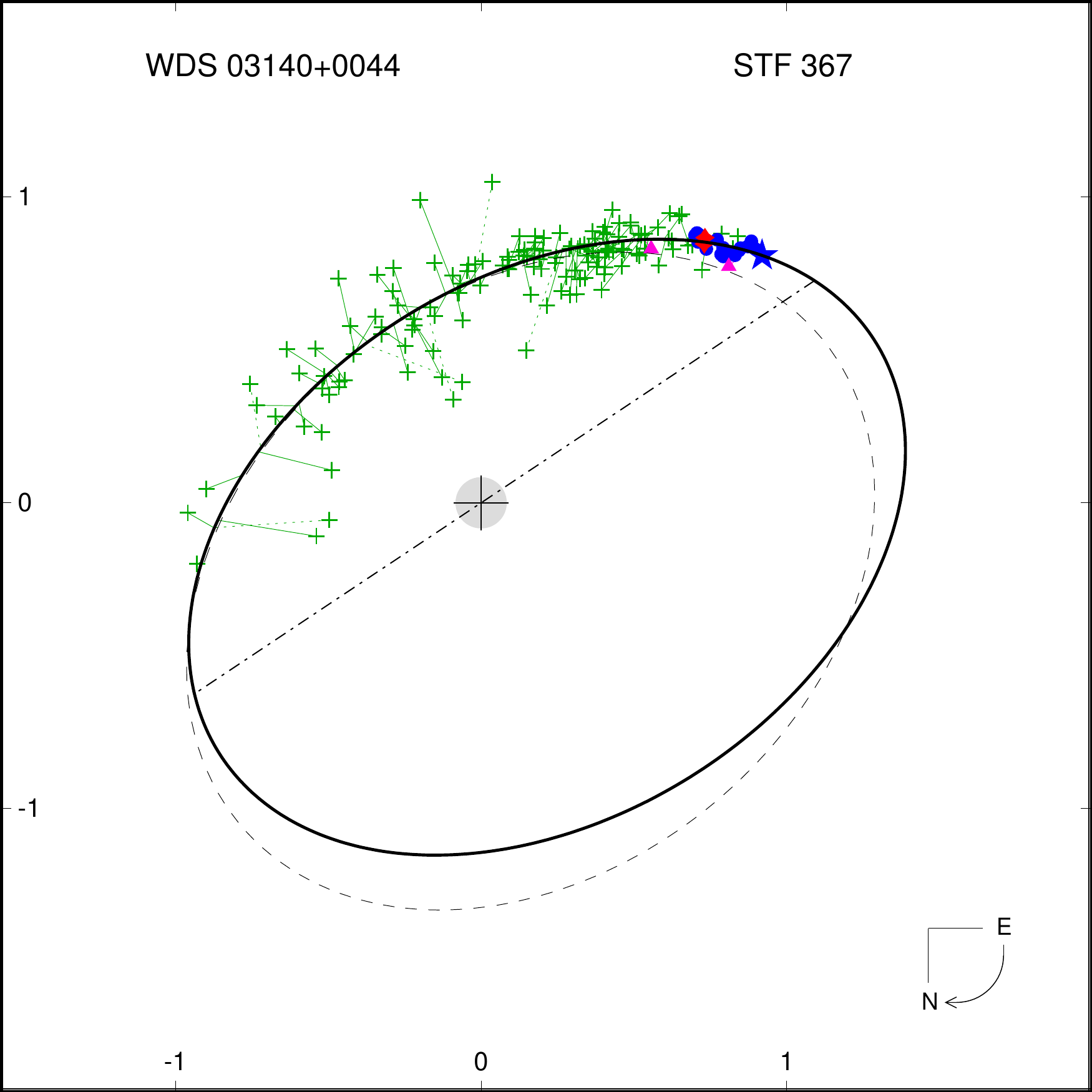}
\end{center}
\caption{\small New orbits for the systems listed in Table~\ref{tab:neworb}, 
together with all published data in the WDS database. See \S~\ref{sec:orbits} for a
description of symbols used in this and the following figures. In all of these figures, Figure a. is the upper left corner, b. the upper right, c. the lower left, and d. the lower right.  In this figure: a. HIP 5898; b. HIP 7916; c. HIP 12204; d. HIP 15058.
\label{fig:neworb1} }
\end{figure}

\begin{figure}[p]
\begin{center}
\epsscale{1.00}
\plottwo{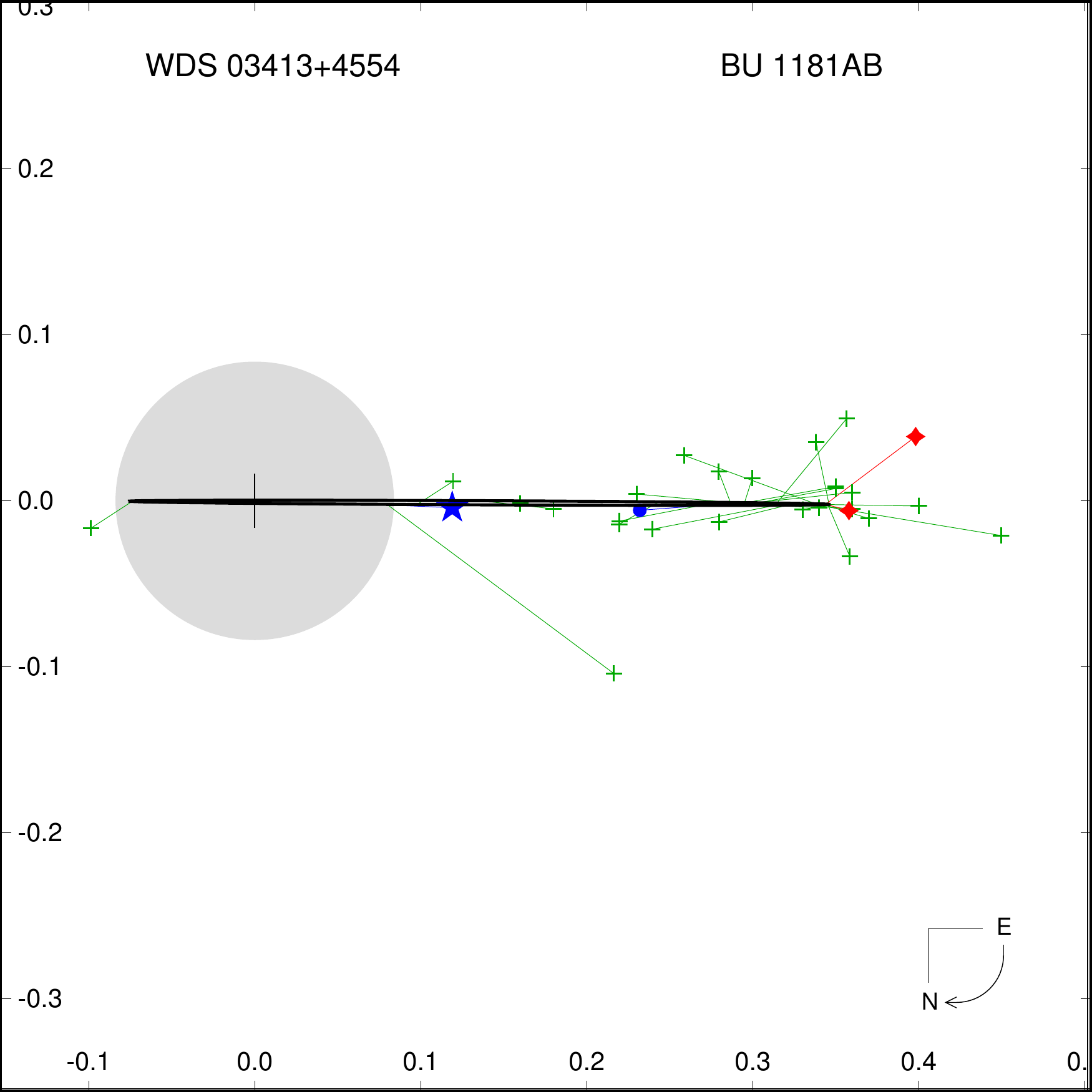}{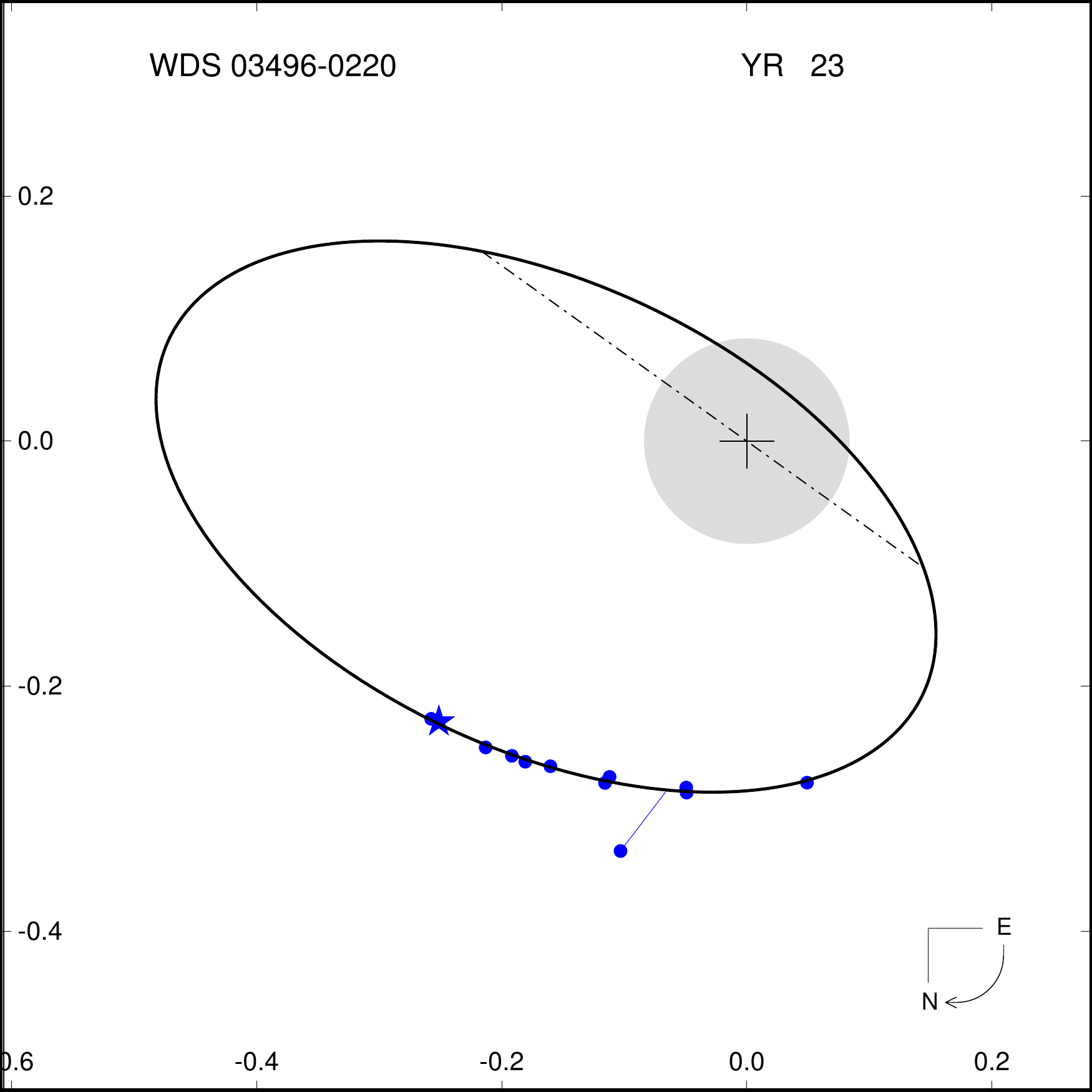}
\vskip 0.05in
\plottwo{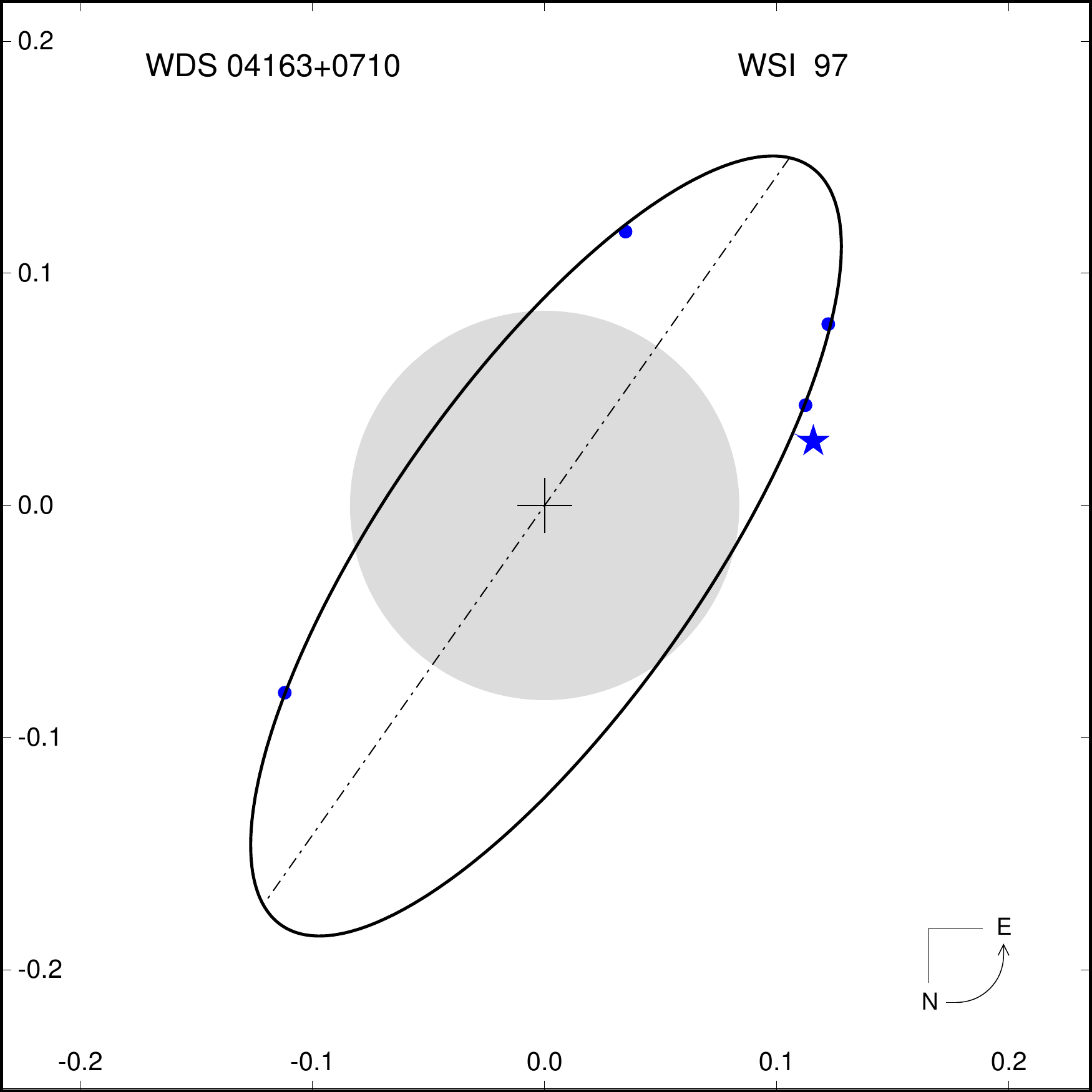}{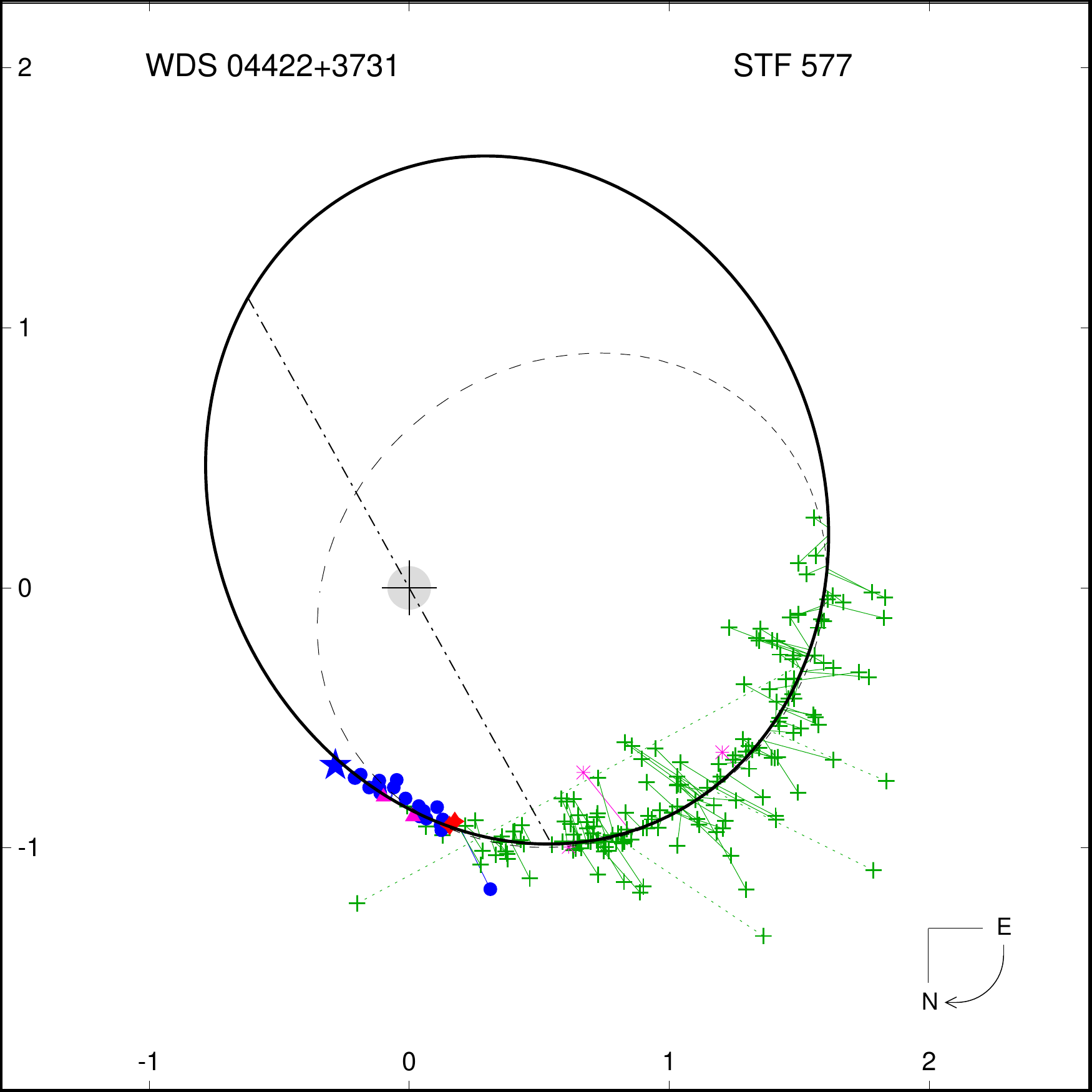}
\end{center}
\caption{\small New orbits for the systems listed in Table~\ref{tab:neworb} (continued).  In this figure: a. HIP 17217; b. HIP 17895; c. HIP 19911; d. HIP 21878.
\label{fig:neworb2} }
\end{figure}

\begin{figure}[p]
\begin{center}
\epsscale{0.85}
\plottwo{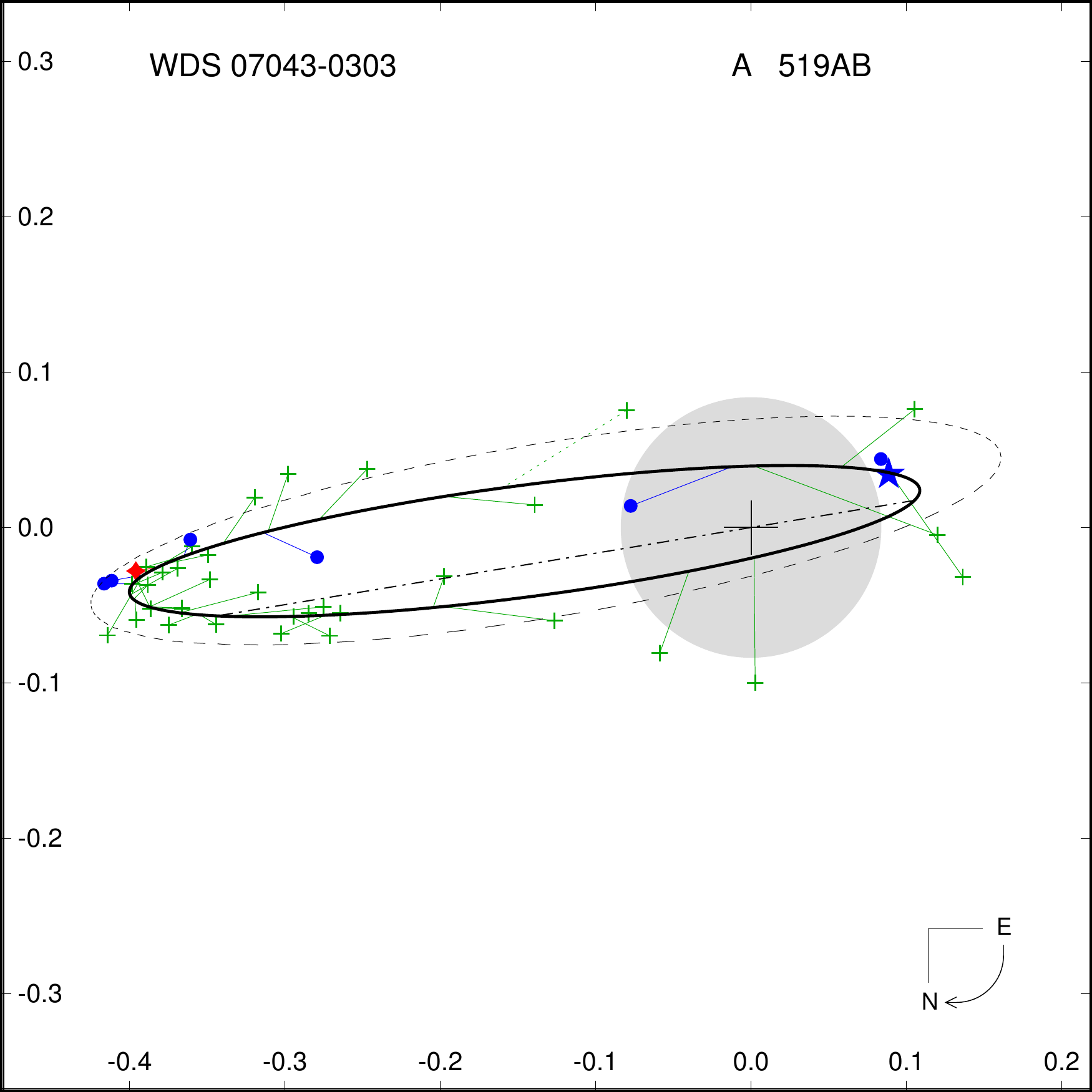}{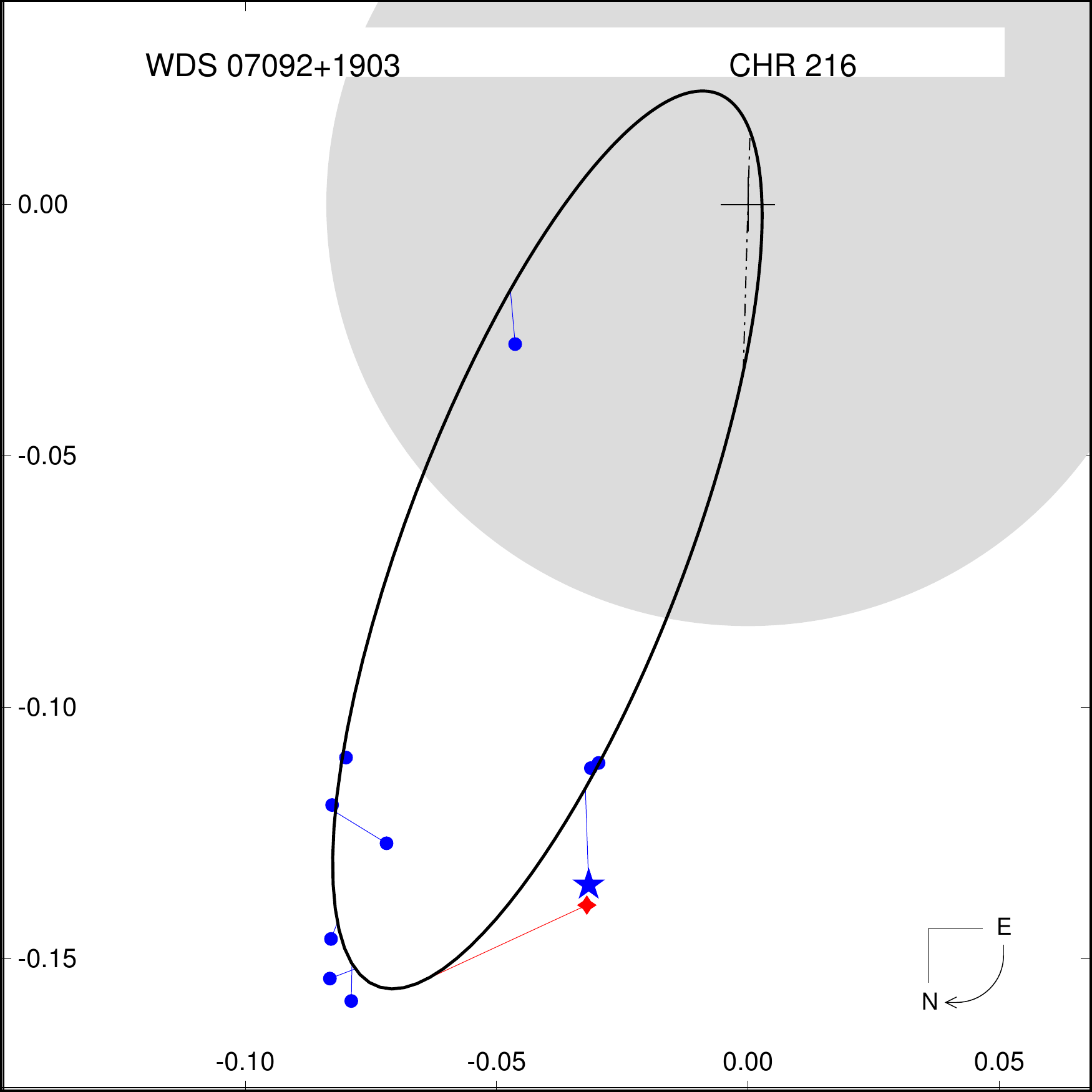}
\vskip 0.05in
\plottwo{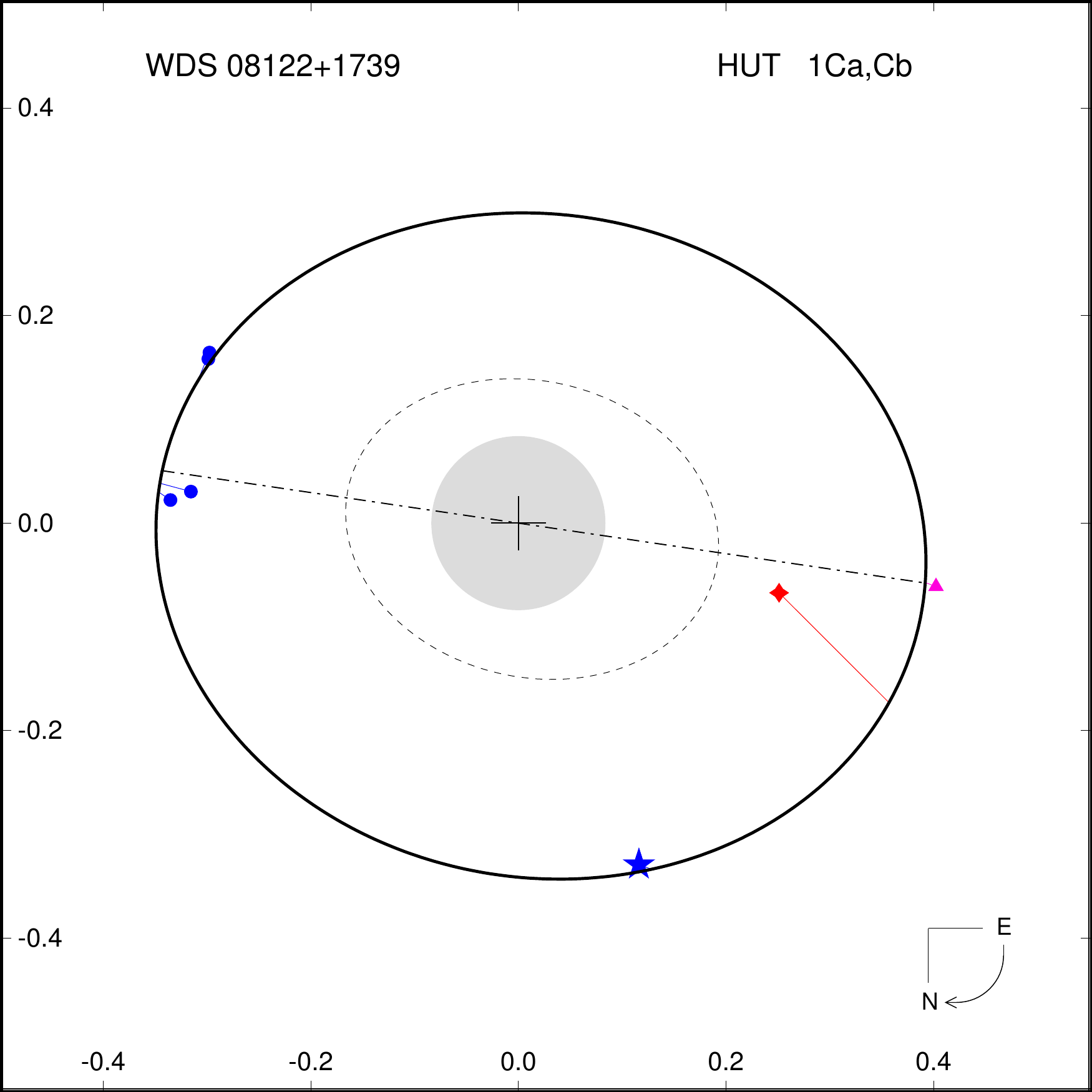}{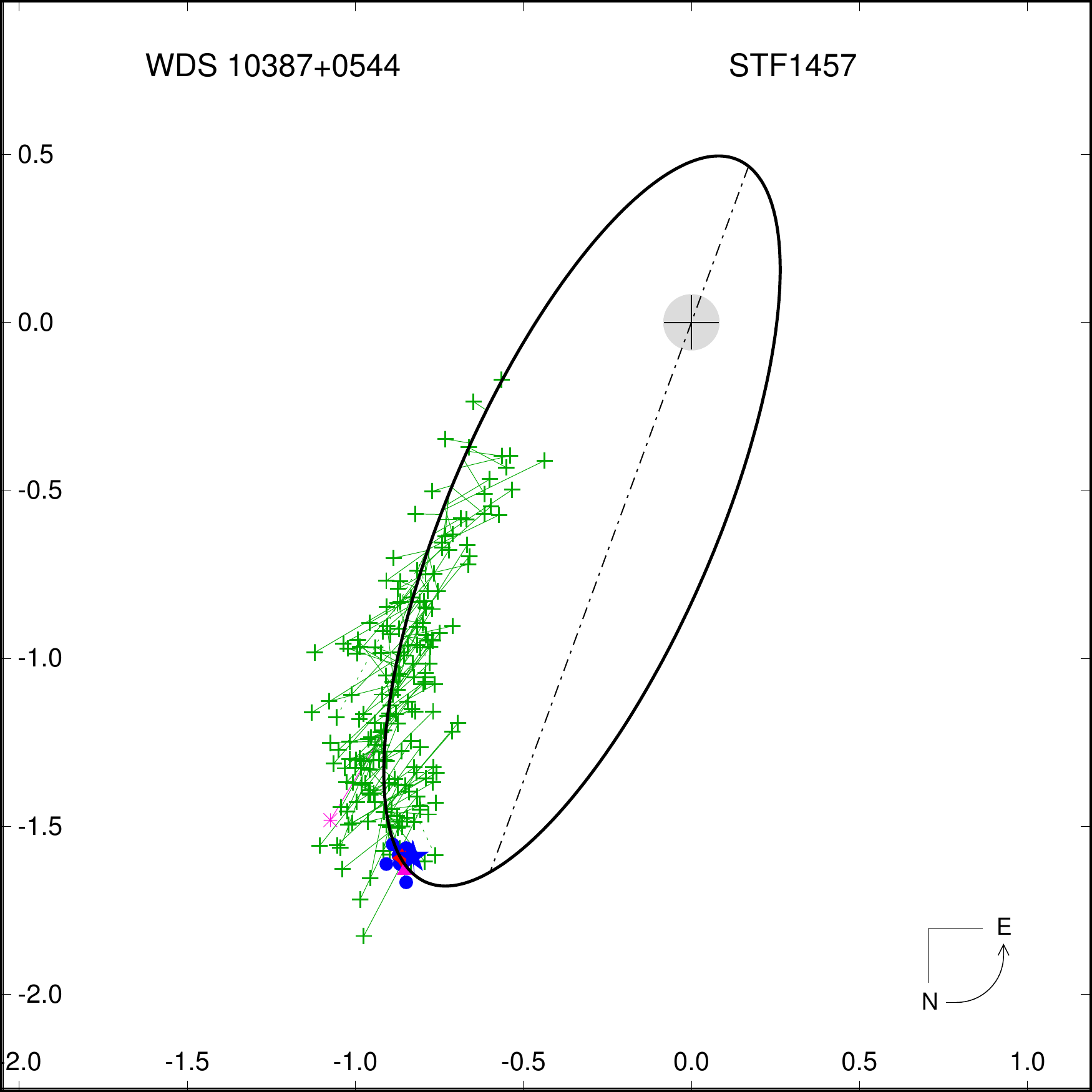}
\end{center}
\caption{\small New orbits for the systems listed in Table~\ref{tab:neworb} (continued).  In this figure: a. HIP 34110; b. HIP 34524; c. HIP 40167; d. HIP 52097.
\label{fig:neworb3} }
\end{figure}

\begin{figure}[p]
\begin{center}
\epsscale{1.00}
\plottwo{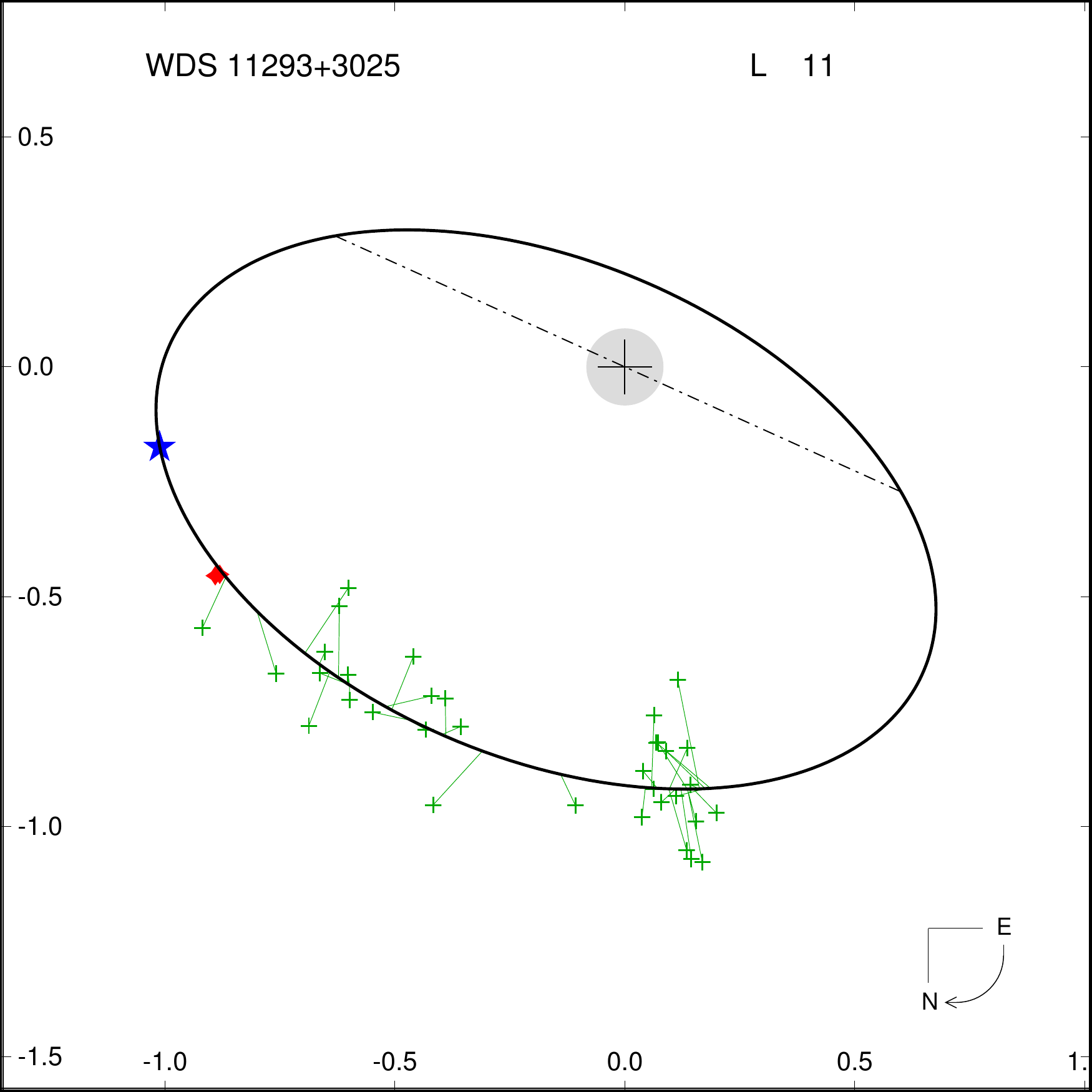}{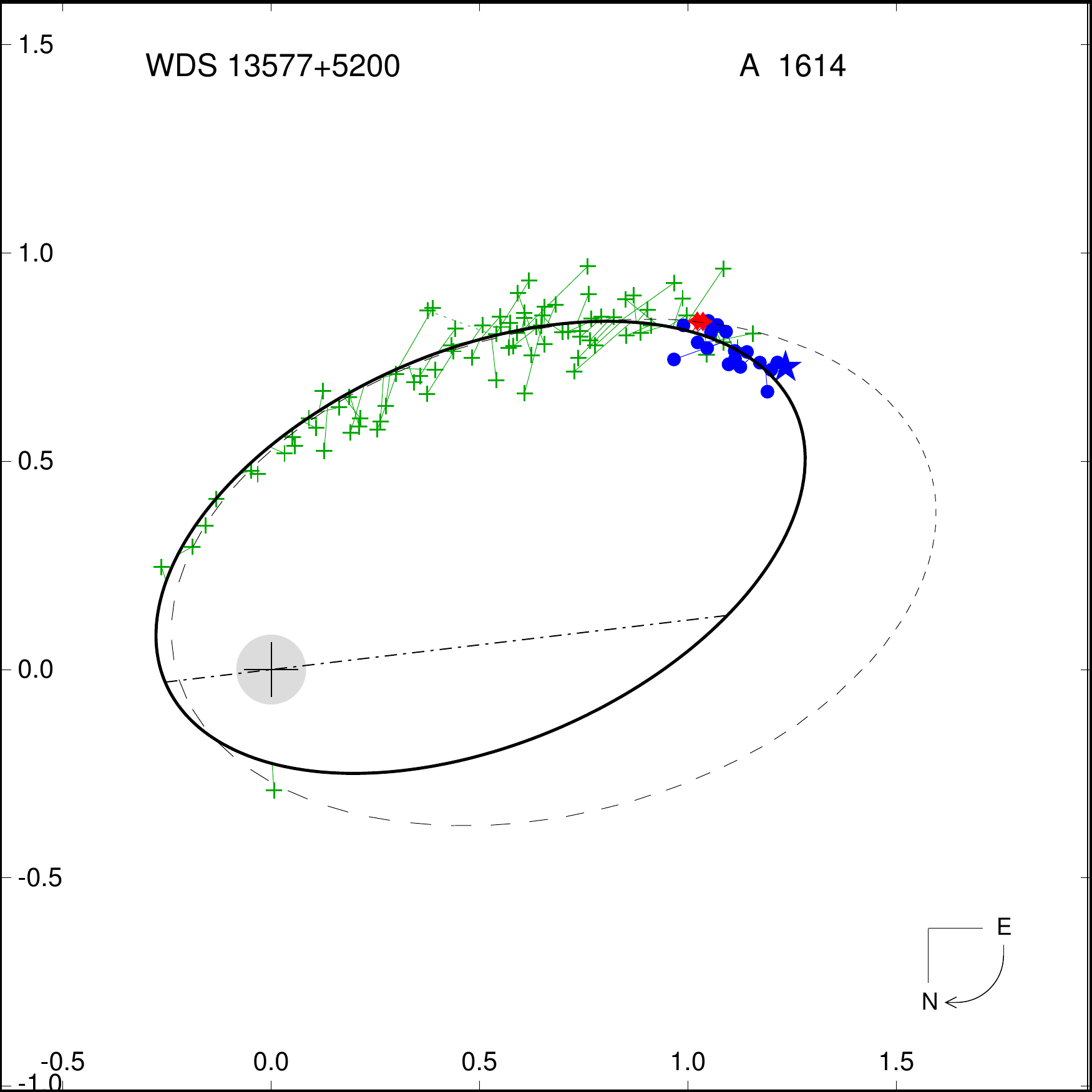}
\vskip 0.05in
\plottwo{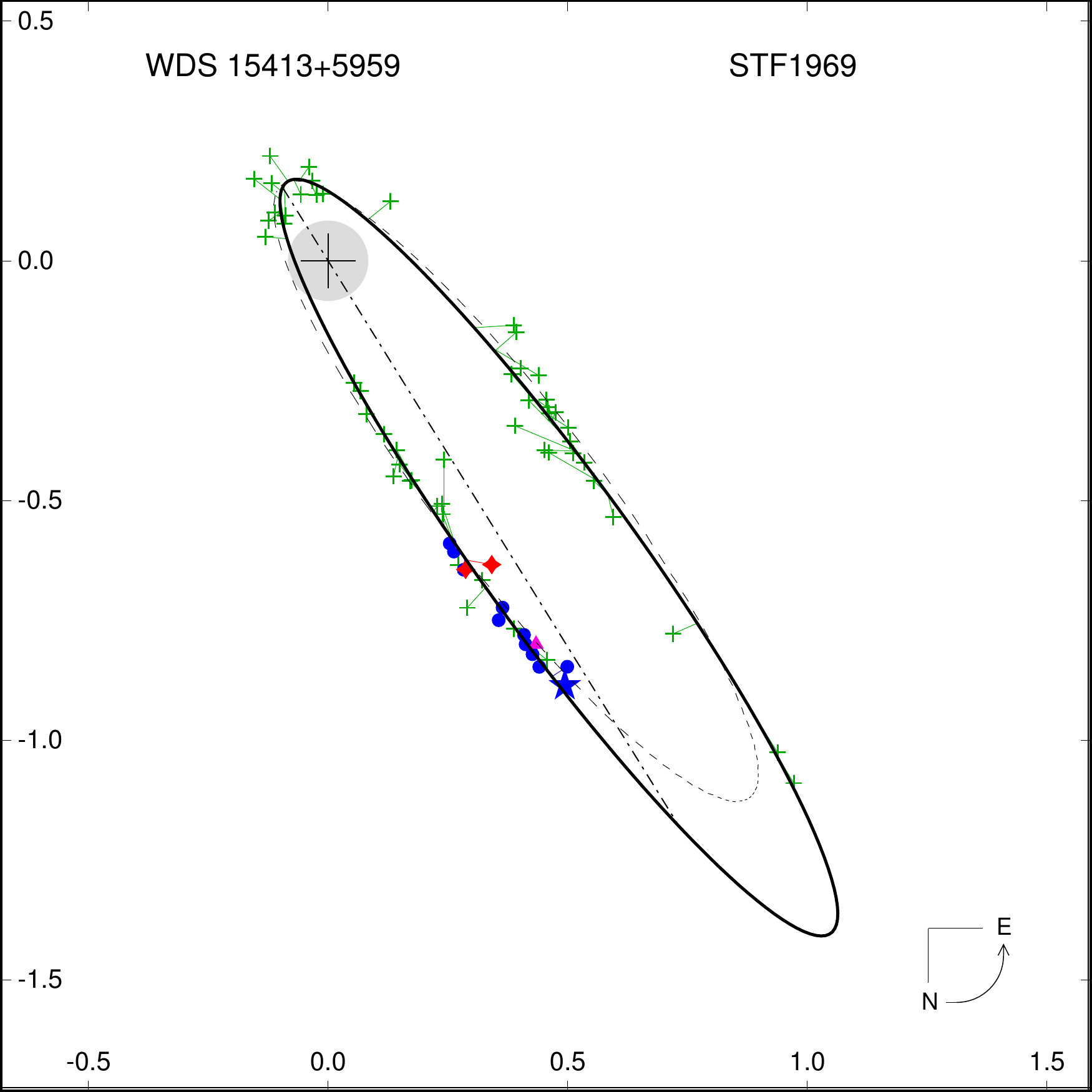}{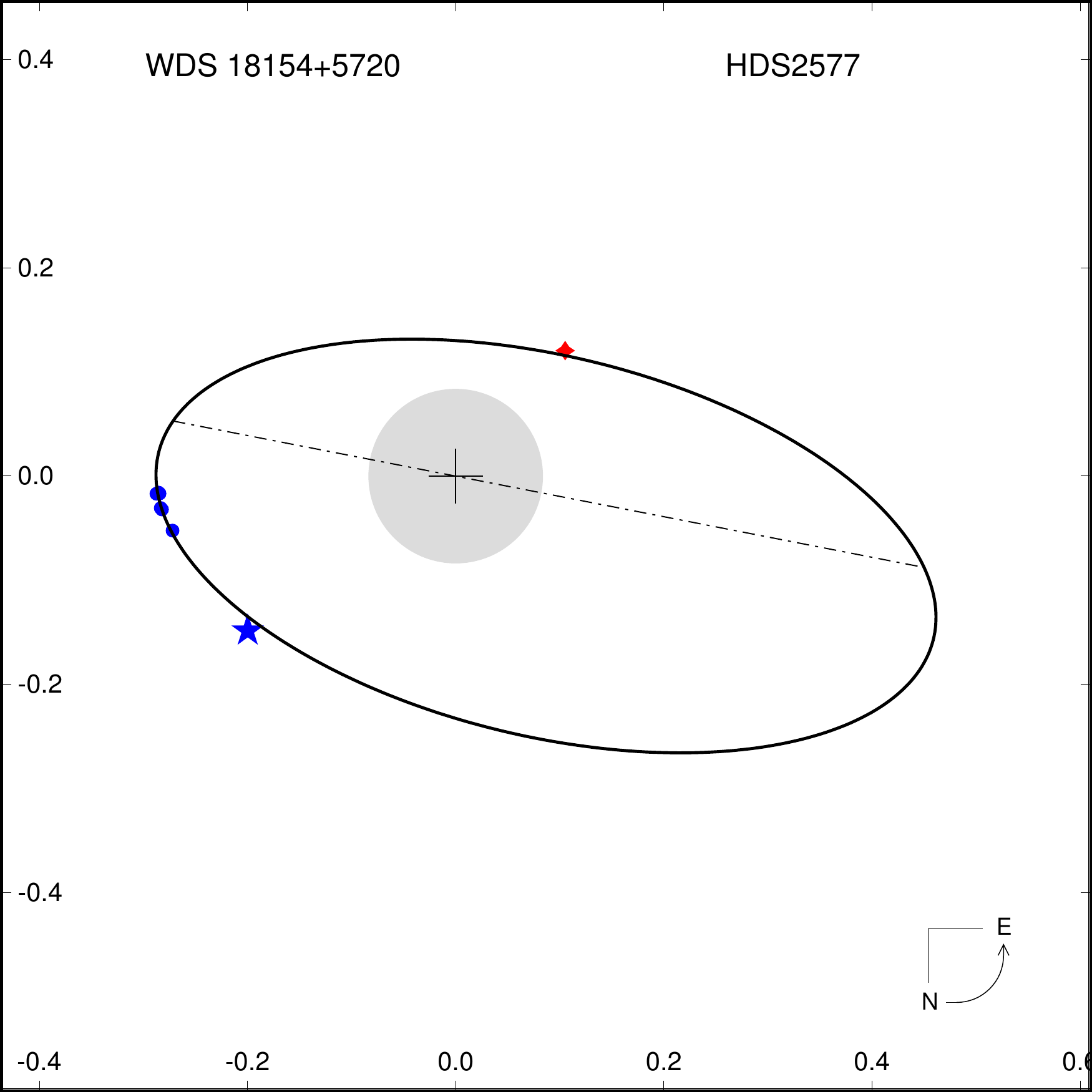}
\end{center}
\caption{\small New orbits for the systems listed in Table~\ref{tab:neworb} (continued).  In this figure: a. HIP 56054; b. HIP 68193; c. HIP 76837; d. HIP 89455.
\label{fig:neworb4} }
\end{figure}

\begin{figure}[!ht]
\begin{center}
\epsscale{1.00}
\plottwo{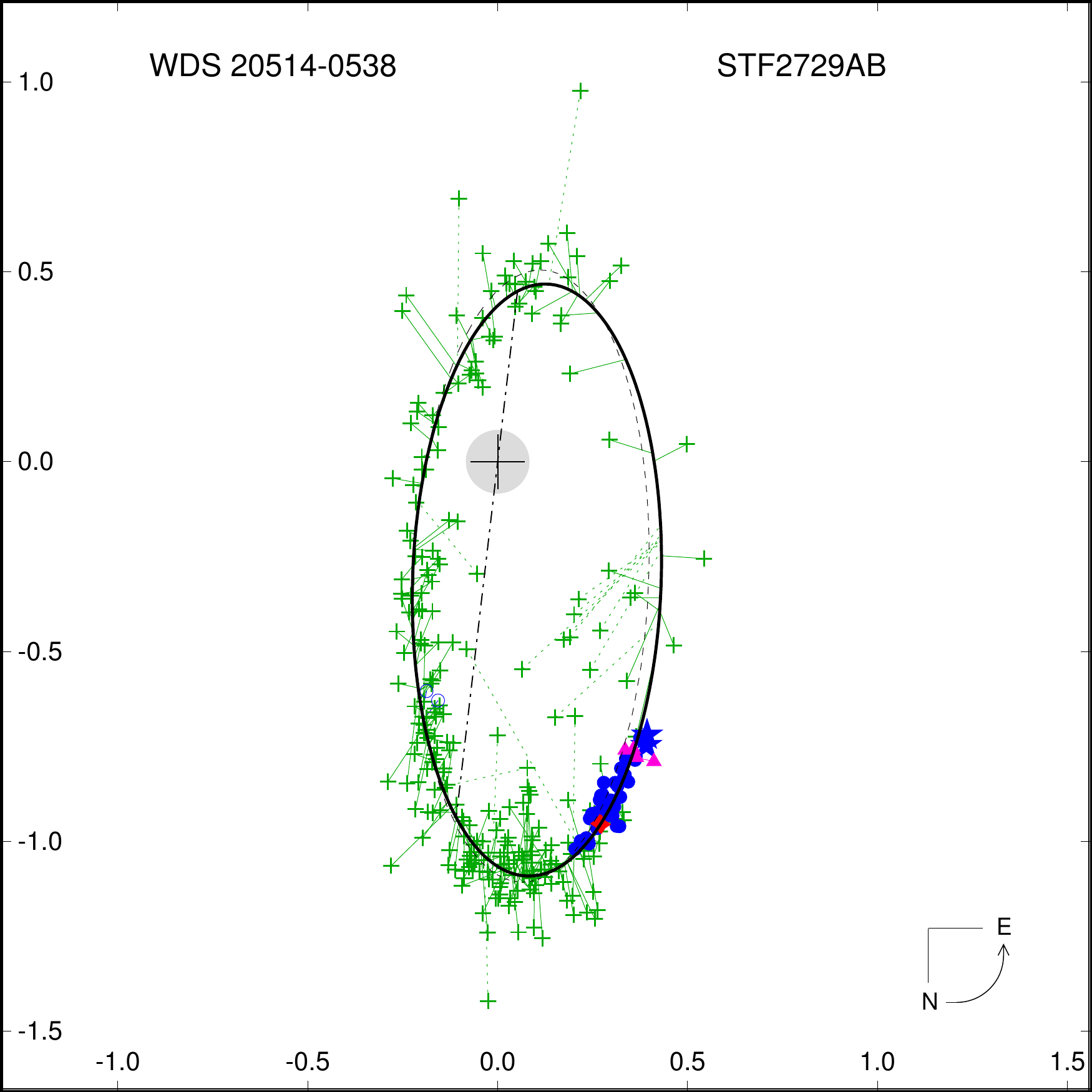}{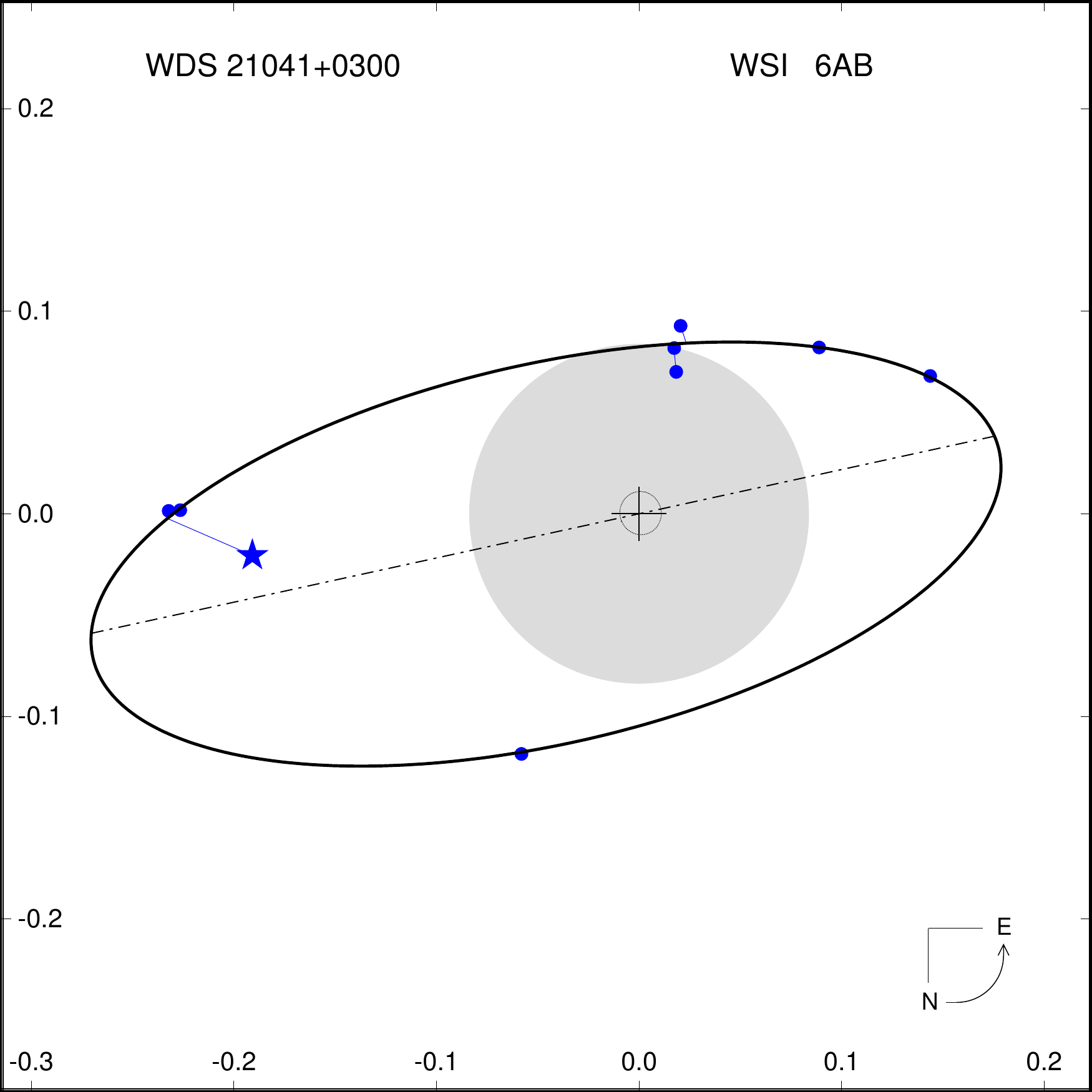}
\vskip 0.05in
\plottwo{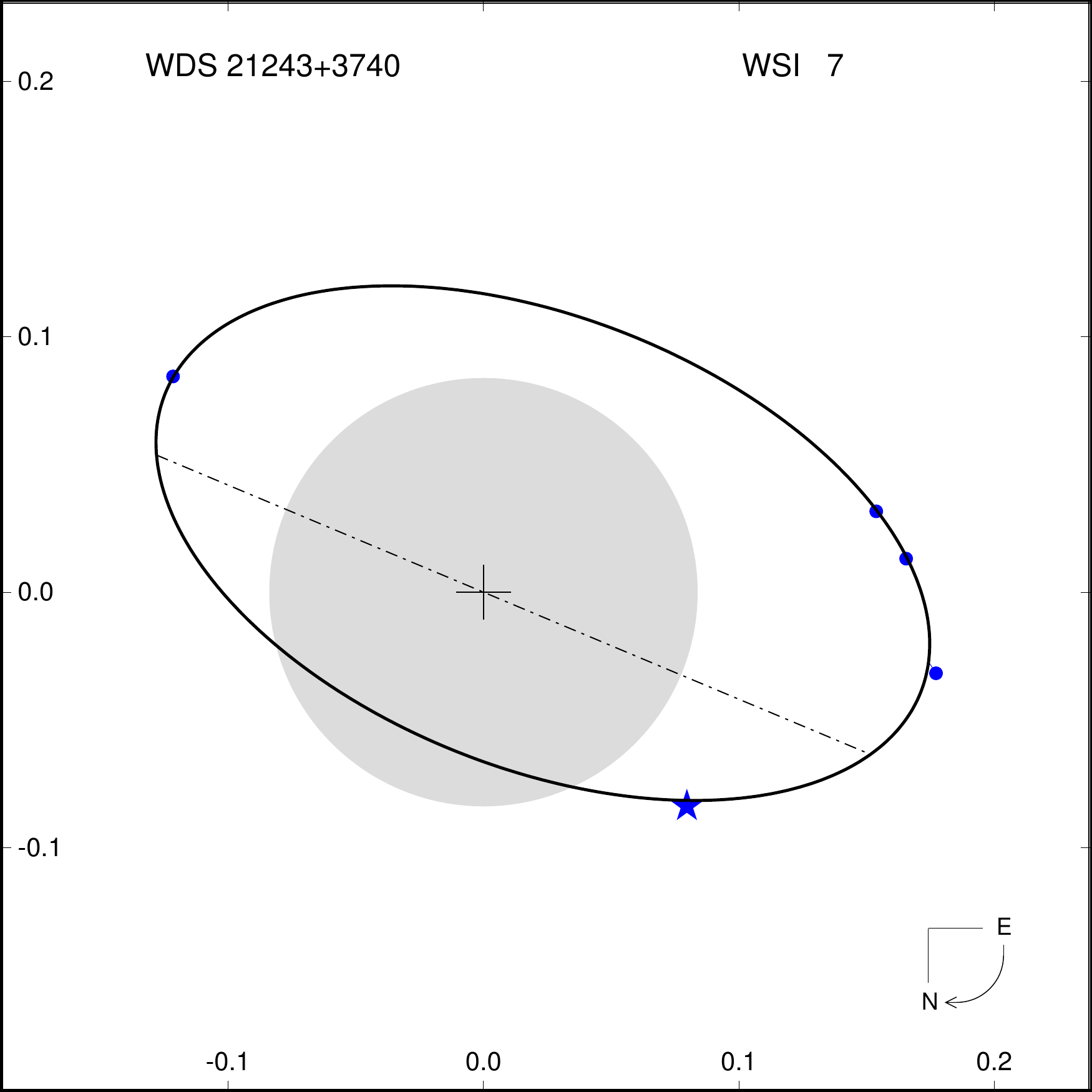}{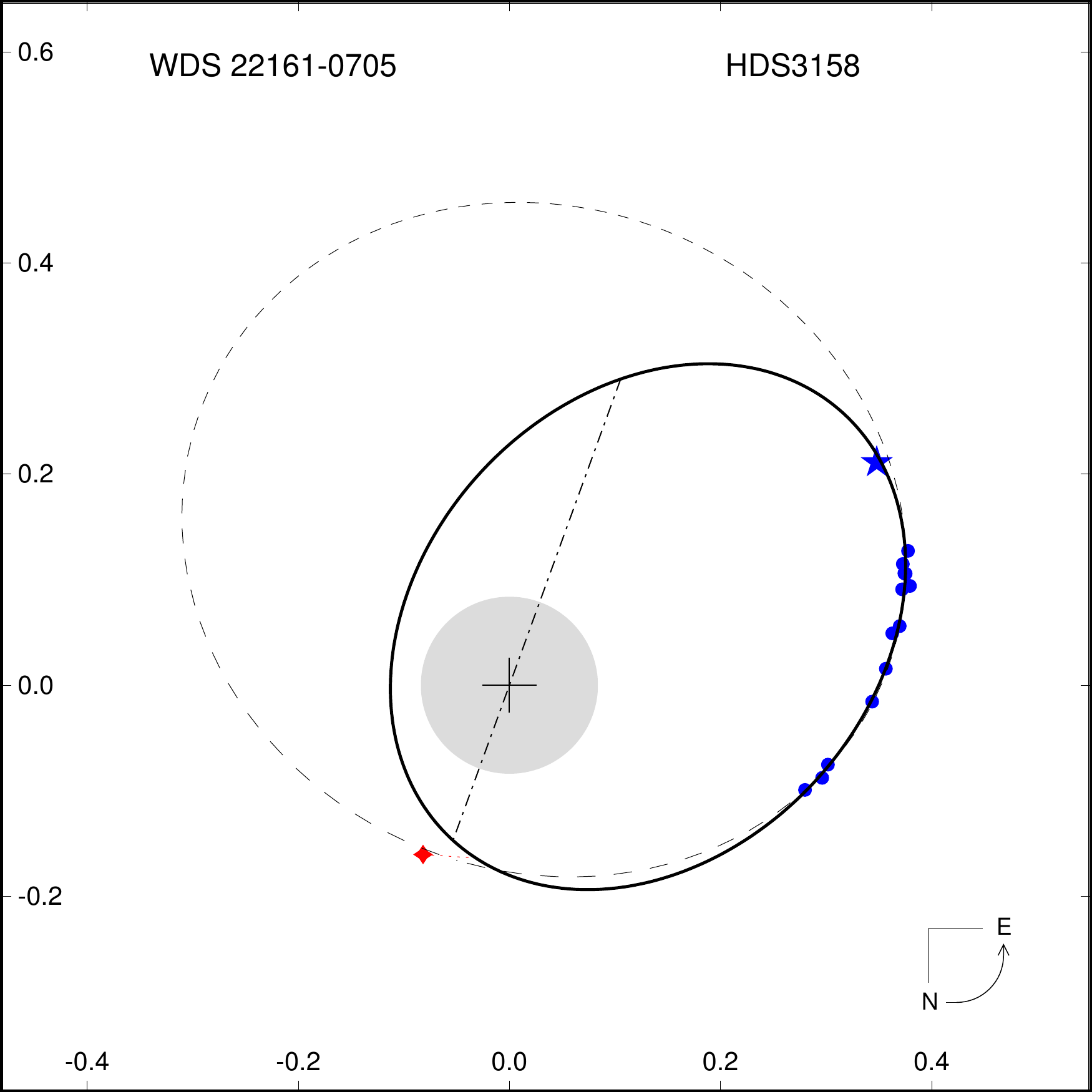}
\end{center}
\caption{\small New orbits for the systems listed in Table~\ref{tab:neworb} (continued).  In this figure: a. HIP 102945; b. HIP 103987; c. HIP 105676; d. HIP 109951.
\label{fig:neworb5} }
\end{figure}

New orbits  are presented  in Table~\ref{tab:neworb}.  In  this Table, the  pairs are
identified  by  their WDS and HIP1 numbers, followed by their orbital  elements, giving the period  $P$ in years,
the  semimajor  axis  $a$  in  arcseconds,  the  inclination  $i$  and
longitude of the node $\Omega$, both in degrees, the epoch of the most
recent   periatron   passage  $T_\circ$   in   Besselian  years,   the
eccentricity $e$ and the longitude of periastron $\omega$ in degrees.
Following this  is the orbit grade \citep[see][for  details]{VB6} as an
evaluation  of  the  orbit.   For  those  pairs  with  previous  orbit
solutions we provide a reference to the previous ``best'' orbit.  Figures~\ref{fig:neworb1} to~\ref{fig:neworb5} show
the new orbital solutions, plotted  with all published data in the WDS
database.   In each  of  these figures,  micrometric observations  are
indicated by  plus signs, high resolution measures  by filled circles,
Robo-AO observations as filled stars, and {\em Hipparcos} observations
as  filled diamonds.   An  O$-$C  line connects  each  measure to  its
predicted position along the new  orbit (shown as a thick solid line).
The shaded circle centered on the primary indicates the resolution
limit of the P60. A dot-dashed line  indicates the line of nodes, 
and  a curved arrow in the  lower right  corner of  each  figure indicates  
the direction of orbital motion. Previous published orbits are shown as  
dashed ellipses; references to each of the published orbits are given in the notes to Table~\ref{tab:neworb}.  

Orbital  elements were  determined  with {\it  orbgrid10}, the  latest
version    of    the   venerable    orbit    reduction   package    of
\citet{Hartkopf1989} which utilizes a three dimensional ($P$, $T$, and
$e$)  grid search  of  variable  grid step  sizes  from three  initial
estimates.  It  has been  modified to include  weighting methodologies
determined in  the production of  the 5$^{\it th}$ Catalog  of Orbital
Elements of Visual Binary Stars \citep{Hartkopf2001}.  Some orbits use
new   speckle    data   from    the    SOAR   telescope
\citep{Tokovinin2014b}.  

  The rest  of this Section gives notes on  some pairs with new
orbits.

HIP 12204 (=WDS J02371$-$1112 = HU 1216AB) has only a preliminary orbit of grade 5, hence the formal errors of the elements are very large.

HIP~17895 (=WDS  J03496$-$0220 = YR~23):  This is  the first calculated orbit of
this pair  announced in \citet{Horch2002b} based  on observations with
the  WIYN  telescope.  The  2004  measure,  obtained  with the  26~inch
telescope in Washington, D.C.~\citep{Mason2006}, is  off a bit, which is not
surprising given the aperture.

HIP~34524 (=WDS J07092$+$1903 = CHR~216) was  resolved by lunar
  occultations  by   \citet{Africano1975}  and  has   several  speckle
  measurements.  Robo-AO finds it at  nearly the same position as {\em
    Hipparcos}  (the quadrant  of our  measure is  established  by the automatic data reduction co-added image),  so the binary has made nearly one full  revolution since
  1991.25.   It  was not  resolved  at  SOAR  in 2009--2011  ($\rho  <
  0\farcs03$),  but measured  in 2012.9,  2013.05, and  2014.06 (these
  data  are  used  in  the  orbit calculation).   The  orbit  is  very
  eccentric, but measurements do not constrain the eccentricity, so it
  was fixed  to 0.94  -- a provisional  value that gives  the expected
  mass sum of 1.8~${\cal M}_\odot$.   The orbit should be followed by
  spectroscopy.   Unfortunately, the  periastron passage  in  2010 was
  missed, we have to wait for the next one in 2038.  

HIP~40167 (=WDS J08122$+$1739 = HUT~1Ca,Cb): This is the first calculated
orbit of this pair. The existence of the Ca,Cb pair was inferred and
the first astrometric orbit calculated by \citet{Heintz1996b}.

HIP~103987 (=WDS J21041$+$0300 = WSI~6): This is the first calculated orbit of this pair first
split  by \citet{Mason2001}  with the  Struve 2.1~m  telescope  in an
investigation  of  \textit{Hipparcos}  acceleration (G-type  solution)
pairs.

HIP~105676 (=WDS J21243$+$3740 = WSI~7): Like HIP~103987 above,
this is the first calculated orbit of this pair also first resolved with the 
same telescope on the same project.

\section{Summary}
\label{sec:sum}

A survey of high-resolution   imaging  with  Robo-AO   has  advanced  our
knowledge of the multiplicity of solar-type stars in several ways.

The binarity of the  wide secondary components turned out to be
comparable to the frequency  of sub-systems in the primary components.
This   overturns  the   established   paradigm  about   multiple-star
architecture,  where  sub-systems were  thought  to be  preferentially
associated with  the primary, most  massive star in the  system.  This
traditional  view   can  be  traced  to   the  previous  technological
restrictions allowing to study  only bright stars by spectroscopy or
high  angular  resolution, and matched  the idea  that
multiple  stars experience  chaotic dynamical  interactions  where the
least massive components are ejected  onto distant orbits or leave the
system altogether.  This new paradigm of dynamical evolution of multiple star systems, where wide secondary  components retain companions at the same frequency as the primary, will require more exploration to determine how such systems develop.

Quite unexpectedly, we discovered  that sub-systems in the primary and
secondary   components   of    wide   binaries   are   often   present
simultaneously, i.e.  these systems  are in fact 2+2 quadruples.  This
new  result means that  2+2 quadruples  are relatively  frequent. Note
that  in  the  25-pc  sample  of \citet{Raghavan2010},  9  out  of  11
quadruples  have the  2+2 architecture  and only  two are  of  the 3+1
``planetary''  configuration; this result, based in small-number statistics,  is  now
supported  by the  much larger  FG-67 sample  studied with Robo-AO.   The high
frequency  of  2+2  quadruples  challenges  the  paradigm  of  chaotic
$N$-body   dynamics  (which   can  produce   such   architecture  only
exceptionally) and calls for  exploration of other formation scenarios
for multiple stars \citep[e.g.][]{Whitworth2001}.

The high  dynamic range  of the  Robo-AO imaging  enabled discovery  of 17
certain  and  2  possible   tertiary  components  to  close  binaries,
converting them  into triple  or higher-order systems.   Statistics of
such triples  reveal some interesting  details, such as the  prevalence of
outer separations  on the  order of 500~AU.   This result is  not yet
formally  significant,  but  it  highlights  the  potential  of  large
samples.    Historically  small  binary   samples  allowed   only  the
first-order description of binary  statistics by smooth functions with
a  few parameters \citep{Duchene2013}.   Modern large  and homogeneous
samples begin  to reveal new  details in these distributions,  such as
twin binaries  with identical masses  \citep{Lucy2006}, bimodal period
distribution in  Hyades \citep{Griffin2012}, and,  possibly, preferred
separations  of tertiary  components seen  here.  These  findings will
advance our understanding of multiple-star formation.

The  large data  set  that resulted  from  this survey  enabled us  to
compute  first visual  orbits for  9  pairs and  to update  (sometimes
dramatically)  the  existing orbits  for  11  more.  Accumulation  and
improvement of the  visual-orbit data will lead to  a better knowledge
of stellar masses and to the statistical analysis of orbital elements.
It  will also  enable dynamical  study of  visual  binaries presenting
special interest.

\acknowledgments 

We acknowledge the input of the referee that read through this lengthy paper and gave us comments to improve it.  The Robo-AO system is supported by collaborating partner institutions, the California Institute of Technology and the Inter-University Centre for Astronomy and Astrophysics, and by the National Science Foundation under Grant Nos. AST-0906060, AST-0960343, and AST-1207891, by the Mount Cuba Astronomical Foundation, and by a gift from Samuel Oschin. We are grateful to the Palomar Observatory staff for their ongoing support of Robo-AO on the P60, particularly S. Kunsman, M. Doyle, J. Henning, R. Walters, G. Van Idsinga, B. Baker, K. Dunscombe and D. Roderick.

A portion of the research in this paper was carried out at the Jet Propulsion Laboratory, California Institute of Technology, under a contract with the National Aeronautics and Space Administration. 

C.B. acknowledges support from the Alfred P. Sloan Foundation.

This work  used the  SIMBAD service operated  by Centre  des Donn\'ees
Stellaires  (Strasbourg, France),  bibliographic  references from  the
Astrophysics Data System maintained  by SAO/NASA, data products of the
Two Micron All-Sky Survey  (2MASS), the Washington Double Star Catalog
maintained    at   USNO.    

{\it Facilities:} \facility{PO:1.5m (Robo-AO)}

\appendix

\section{Appendix:  Notes on new binary star systems}\
\label{appendix}

HIP~2292B = HIP~2350 = RAO 1BC is resolved at 0\farcs5, with several
Robo-AO measures and a confirmation  with the Palomar 200~inch (5~m) telescope (P200). 
It remained fixed during
1\,yr, despite the proper  motion (PM) of 0\farcs25 yr$^{-1}$. This star
also   hosts    an   exoplanet   system    with   $P=3.44$\,d.    The
common-proper-motion (CPM)  component A is at 839\arcsec  ~from B.  We
discuss this multiple system in a separate paper.

HIP~2717B  =  BD+61~119 ($V=8.98$,  F8V,  X-ray  source)  has two  new
companions at  $\sim$3\arcsec, designated as RAO~2BC  and RAO~2BD. We
see the brighter companion C in the 2MASS image at a similar position,
although it is not listed as  a separate point source in 2MASS. On the
other hand, the fainter star D  is likely optical because the field is
crowded and because a trapezium-like  configuration of BC and BD would
be  dynamically  unstable.   The  CPM  nature  of  AB  \citep{LEP}  is
confirmed by the data from 2MASS and UCAC \citep{Zacharias2013}. 

HIP~3214B  ($V=14.87$)  is  resolved  at  1\farcs6. The  new  pair  is
considered  physical because the  companion is  bright and  because it
remained fixed during  1\,yr. The 2MASS image is  elongated in the E-W
direction.

HIP~3540 AB is  LDS~836 at 55\farcs4 separation. We  resolved B into a
new 1\farcs5 pair RAO~4BC, confirmed with the P200.  The component  A is a
spectroscopic binary  (SB) with $P=11.4$\,d  \citep{Griffin2002}.  The
system is thus a 2+2 quadruple. 

HIP~3795 is  another new quadruple  system. The AB pair BU~495  has an
orbit  with  $P=413$\,yr.   The   CPM  component  C  ($V=10.83^m$)  at
152\arcsec ~from AB  turns out to be a  new 0\farcs67 pair RAO~5Ca,Cb.
We  consider Ca,Cb  physical   because  the  companion  is  bright,  the
separation is small, and the field is not crowded.

HIP~4016B is  accompanied by  the 1\farcs7 pair  CD, at a  distance of
17\arcsec ~from B  (AB = DAL~11 at 41\arcsec ~is  physical). The field is
crowded, so we consider the faint stars C and D to be unrelated to B.  BC is at $(267.2^\circ, 15\farcs66)$ in 2MASS. 

HIP~4878 has faint  stars B and C at  2\farcs9 and 32\farcs2 distance,
respectively (RAO~39).  The optical  nature of C can be established by
its different  position (145\fdg3, 32\farcs4) and its  color in 2MASS.  AB was re-measured  with the P200 one
year later and found at a slightly different position, so it is likely
optical as well, although the field is not crowded.

HIP~5276  is  an  SB  with  $P=49.5$~d~\citep{2014MNRAS.441.2316G},  accompanied by  the newly  resolved  tertiary component
RAO~40 at 6\farcs2.  The physical  nature of the tertiary is confirmed
by the  2MASS  position and by observations with P200.

HIP~5313  is  another triple  system:  Aa,Ab  is  an acceleration  and
spectroscopic binary  with a distant physical component  B (RAO~41) at
26\farcs2, confirmed by its position and color in 2MASS.

HIP~6653 has a  new companion at 1\farcs8 (RAO~6). The  star is on the
Keck exo-planet  program, its radial-velocity  (RV) trend (D.~Fischer,
private communication, 2012) could be caused by the companion.  The companion
B is red, the field has low crowding, so the physical nature of AB is
likely.

HIP~7845 was measured as a test target, it is not in the FG-67 sample.
The  faint   companion  C   at  24\arcsec   ~moved
substantially during a year, hence it is optical. 

HIP~11696 is located  in a crowded region of the  sky. We targeted the
component B  and found a faint  star at 15\farcs7  from it (RAO~42BE),
considered optical (just as the other components C and D listed in the
WDS) because of the high crowding, non-hierarchical configuration, and
the 2MASS position of BE.

HIP~12062 has an optical companion RAO~43 at 14\farcs8. The pair has a
different position in 2MASS and the region is crowded. 

HIP~12067 has a new 5\farcs7 companion B (RAO~7), confirmed  by the 2MASS
position, in addition to the inner acceleration and
spectroscopic binary with yet undetermined period.

HIP~12189 and  HIP~12184 form a 38\arcsec ~physical  pair STFA~5 where
each component is, in turn, a close binary: A is SB ($P=1.1$\,d), B is
resolved here  into a 0\farcs53  pair RAO~8, confirmed next  year with
Robo-AO and P200.  Moreover, an exoplanet with $P=335$\,d around B was
announced. We discuss this multiple system in a separate paper.

HIP~12685 has a companion at  21\farcs7 (RAO~44) which is also present
in 2MASS at approximately the same position. However, the PM is small, the
field is  crowded, and the color  of B does  not match a dwarf  at the
same distance as A, so we consider B optical.

HIP~12925 has a new companion at 1\farcs9, measured several times with
Robo-AO and  confirmed with the P200. This  is a young  multiple system: the
radial velocity (RV) of  A is variable \citep{Nordstrom2004}, there is
a CPM companion at  494\arcsec~\citep{LEP}, and HIP~12862 at 54\arcmin~
is co-moving.

HIP~13336  contains  the  visual  binary  A~281AB  and  a  companion
RAO~45AC at  15\arcsec, considered  optical because  it is  found at
a different position in 2MASS.

HIP~15329  contains  the  AB pair   (STF~53, $P=113$~yr)  and  a  CPM
companion  C ($V=13.24$) at  107\arcsec~found  by \citep{LEP}.   Here C
is resolved at  1\farcs6 (RAO~10Ca,Cb),  revealing this system  as a 2+2
quadruple.  We believe  that Ca,Cb  is physical  because of  its small
separation, moderate  crowding, and the red  color of Cb  ($\Delta i =
4.0$ mag, $\Delta z = 2.8$ mag).

HIP~16329 Ba,Bb = RAO~11Ba,Bb (also known as J~207) is a spectroscopic
binary  in  the  Hyades  cluster  with a  period  of  about  30.74~yr
\citep{Griffin2012}.  Assuming the mass of  the K7V component Ba to be
0.7~${\cal M}_\odot$, the spectroscopic  orbit implies a minimum mass
of 0.3~${\cal M}_\odot$  for Bb and a semi-major  axis of 0\farcs265.
The pair Ba,Bb  was resolved by Robo-AO three  times at 0\farcs16 with
$\Delta i  = 0.87^m$, $\Delta  r = 0.97^m$,  and $\Delta z  = 0.52^m$.
This corresponds  to a mass for Bb of $\sim$0.6~${\cal  M}_\odot$. We can't
help noting  that Bb appears redder  than Ba, and that  its lines were
not detected in  the spectrum; it could itself by a  close binary or a
fast rotator.  This  pair is a secondary component  in the wide binary
STF~399AB at 20\arcsec ~separation, also measured here (despite its
WDS designation, AB  was first resolved by W.~Herschel  in 1782, not by
W.~Struve).  The  primary component A~=~vB~3 is itself  a close binary
with  astrometric acceleration  \citep{MK05}, although  Griffin states
that RV(A) is constant; it is not resolved by Robo-AO.

HIP~17022 has an astrometric  orbit with $P=3.06$~yr.  This binary is
too  close  to  be  resolved  with Robo-AO, but  we  found  another  faint
star at 1\farcs7 (RAO~47) which is confirmed as physical with the P200.

HIP~17217  is a  triple system  with inner  binary BU~1181  and  a new
component at 4\farcs8 (RAO~48AC) confirmed as physical by its presence
in the 2MASS image.

HIP~17336 has a  new companion at 10\farcs4 which is  not found in the
2MASS point-source catalog, but is  seen at 270$^\circ$ and 12\arcsec ~from
A in the J-band image.  The difference in positions matches the reflex
PM of A, hence the new companion is optical.

HIP~18218  is the binary  A~1293AB. The  new  companion C  at 26\arcsec  ~is
optical, as evidenced by its position in 2MASS
and high crowding.

HIP~18719  is an  SB in  the Hyades  \citep{Griffin2012} which  is not
resolved  here. The  faint  star  at 5\farcs9  found  here is  optical
because it is not seen in the 2MASS images.

HIP~19389  has a CPM  companion C  at 61\arcsec~\citep{LEP}, targeted
here.  The  pair CD is  found in 2MASS, which shows that D is  optical to C,  as the
change in its position simply reflects the PM(C).

HIP~21099 is a member of the  Hyades.  The faint 0\farcs9 pair BC seen at 19\arcsec ~from A is optical, as 
evidenced by its position in 2MASS.

HIP~21443 has a new companion B at 5\farcs7, revealed as physical by its
color  and position  in  2MASS  and confirmed
with the P200. However, the PM(A) is  small and the region is crowded, so the
status of the  new companion is not certain.  The  main target is also
an  SB1 with a  period of  2.06~d \citep{2014MNRAS.441.2316G}.

HIP~23396 is the binary HU~445 which deviates from its orbit. The star C
at 38\arcsec ~is  classified as optical by comparing  with its position
in 2MASS.

HIP~24016 is a  double-lined SB with a CPM companion  B = HIP~24005 at
69\arcsec. We  targeted B  and found a  star C at  106\fdg2, 26\farcs2
from it in this crowded field.   The star is at 103.9\fdg1 and 27\farcs32
in 2MASS, hence it is optical.

HIP~25300 is a system consisting of the 1\farcs1 visual binary STF~677AB
and a new  tertiary component C at 6\farcs8,  confirmed as physical by
its  presence  in  the  2MASS  images. Observations  with the P200  also
confirmed C and resolved B into a close pair.

HIP~26444  is only  triple --  a single-lined  SB coupled  to  the CPM
component  B at 229\arcsec  ~distance \citep{LEP}.  We targeted  B and
found two stars around it  at $\sim$4\arcsec. They are not present in
the 2MASS $K_s$-band image and hence are considered optical.

HIP~27067 and HIP~27070 form  the 22\arcsec ~binary STF~775. We measured
another  star C  at 11\arcsec  ~from A  in this  crowded field;  AC is
optical judging by its position in 2MASS.

HIP~27246 is  a new triple. The 11.1-yr  spectroscopic and astrometric
binary was recently resolved by  \citet{Horch2012}. The new component
C  at 10\farcs9 is  confirmed as physical  by its  position and  color in 2MASS.

HIP~31267 is another  triple system consisting of an  SB and the newly
found tertiary component  C, confirmed as physical by  the 2MASS image
and follow-up observations with the P200.  The 53\arcsec ~pair UC~1450AB is
not physical because because the red color of the secondary places it well 
above the main sequence, while  the small  common  PM is  likely a  chance
coincidence.

HIP~33355 is  an SB with  a new companion  at 5\farcs5 which  could be
physical. The  $K_s$-band image in 2MASS is  extended towards a position
angle of 90$^\circ$  and the companion is confirmed  with the P200. Yet, the
small PM of the target makes it difficult to make a firm conclusion on
the companion's status.

HIP~34110 is a triple system with a 5.1-d SB in the visual pair A~619,
which deviates  substantially from its  orbit. The new component  C at
15\farcs7 is optical as shown the 2MASS data.

HIP~35265 has a wide CPM companion B at 927\arcsec~\citep{LEP}, whose
physical nature is not certain, but likely, given the substantial PM. 
We targeted B and measured RAO~58BC at 17\farcs6. This pair has a
different position in 2MASS, hence is optical. 

HIP~38018 has  an optical companion  C at 118\fdg7, 15\farcs4, with  a different
position in 2MASS.

HIP~40298 has  a CPM  companion B at  244\arcsec ~(LDS 2564)  which was
targeted. The pair BC found here is optical. 

HIP~40479 has an optical  companion at 31\arcsec, with different position in
2MASS.

HIP~40918 has  a CPM  companion B =  HIP~40882 at 258\arcsec  ~which is
resolved here into the 2\farcs8 pair RAO~13BC and confirmed by the 2MASS
image. Considering that A itself  is a long-period SB, this is another
2+2 quadruple discovered by this survey.

HIP~41319  has   a  CPM  companion  at   692\arcsec,  likely  physical
considering its large PM. It was targeted here, but the 16\farcs8 pair
RAO~62BC is optical, as revealed by its different position in 2MASS.

HIP~43426  is a  visual triple  with  two companions  at 3\farcs6  and
49\arcsec ~from it. The distant  component C was targeted. We measured
RAO~63CD  at 18\farcs9,  but  this pair  is  optical.

HIP~48273 is a double-lined SB with $P=3.955$~d  \citep{Griffin2003} and
has a CPM companion  B at 1155\arcsec~\citep{LEP}, considered physical
because  of  the substantial  common  PM  and  matching colors.   This
secondary was resolved here at 0\farcs16 (RAO~90Ba,Bb), converting the
system  into a  quadruple.  The pair  Ba,Bb  was measured  at SOAR  on
2013.13  at similar  position.

HIP~49638 is an  astrometric binary with $P=4.2$~yr and an estimated
semi-major  axis of  0\farcs05. We  detect the new
companion RAO~15 at 0\farcs5, converting this into a triple system. 

HIP~69160  is a  double-lined SB  with $P=8.4$~yr  (D.~Latham, private
communication, 2012)  and  expected semi-major  axis  of  0\farcs09.  It  is
resolved here at 0\farcs12.

HIP~69322 has a CPM component B ($V=13.92$) at 606\arcsec~\citep{LEP},
which is resolved twice into a new 0\farcs58 pair RAO~17BC.

HIP~71843 has a CPM component  B at 216\arcsec, $V=14.9$. We measured a
pair  BC at 17\farcs3  twice. Its  position in  the 2MASS  is similar,  but C is  ``bluer'' than B, hence it could be a white dwarf, 
although more likely it is a background star, considering the small PM of the system.

HIP~75676 is a 40\arcsec ~physical binary KU~108 known since 1893. Both
its  components   are  SBs  (D.~Latham,   private  communication, 2012).  In
addition, we  found the new 0\farcs4  pair BC, making  this a quintuple
system.

HIP~79607 is  a known quintuple system.  The  component E (STF~2032AE,
E=HIP~79551=GJ~615.2C) is  resolved here at 0\farcs4 (but  not for the
first  time: Ea,Eb=YSC~152).   Another  star F  at  32\arcsec ~from  E
(RAO~65) appears optical; it is not found in 2MASS.

HIP~79629  has a  CPM  component  B at  107\arcsec,  $V=14.2$.  It  is
resolved here into a 0\farcs14  pair of equal stars, converting binary
into a triple.

HIP~81608 has a CPM component B at 179\arcsec. The 27\arcsec ~pair
RAO~82BC is optical,  it has a different position  in 2MASS.

HIP~85042  has   a  CPM  companion   B  at  49\arcsec   ~according  to
\citet{Raghavan2010}.  The  secondary is  resolved  here at  0\farcs75
(RAO~19BC), while A itself is single.

HIP~86642 is a double-lined SB (D.~Latham, private communication, 2012) with
a newly  discovered tertiary component  at 2\farcs2.  The  2MASS image
does not resolve this pair, but  it is confirmed with the P200, where the
inner binary was also resolved.

HIP~89207 is a 2\arcsec ~binary  AB (A~2260).  Another star C found here
at 17\arcsec  ~is likely optical, as  the field is  very crowded.

HIP~91120 has a CPM companion B at 621\arcsec, $V=11.14$. It is
resolved here at 0\farcs14 (RAO~83Ba,Bb).

HIP~94540 is a visual triple  system where the 1\arcsec ~pair BU~975BC
is  located at 33\farcs6 from A.  However,  it was outside  the Robo-AO field,
and we  measured only the known  optical pair AD at  15\farcs9 and the
new companion E at 2\farcs9 (RAO~84AE).  The star E is not seen in the
the $K_s$-band 2MASS image and is presumed to be optical, until proven
otherwise.

HIP~94666 is an SB for which  we find a distant companion at 3\farcs6,
confirmed in the 2MASS image.

HIP~94905 is a double-lined  SB with $P=5.38$~d. The companion RAO~67
at 6\farcs9  has the same  position in  2MASS.  However, the  field is
extremely  crowded and the  PM(A) is  small, so  the new  companion is
considered to be likely optical.

HIP~95309  is similar to  the previous  case: we  find a  new 5\farcs1
companion to  the SB in  a very crowded  field. However, the  PM(A) is
0\farcs2~yr$^{-1}$,  so comparison with  the 2MASS position confirms the physical nature of RAO~68.

HIP~95769   is   a   spectroscopic   and   astrometric   binary   with
$P=2.26$~yr. The  companion B at 9\farcs9 is optical;  it has a different
position in 2MASS and the field is crowded.

HIP~96395  is  an SB  converted  into  a  triple by  establishing  the
physical  nature  of  the  10\arcsec ~companion  noted  previously  by 
\citet{Fuhrmann2004}.   Its position in 2MASS  is same  as measured  here, 
its  color matches  a main-sequence dwarf.
 
HIP~97222 is  the binary  STF~2576.  It forms  a quadruple  or quintuple
system   with  HIP~97295   at   792\arcsec,  which   is  the   primary
\citep{Raghavan2010}.  The  field   is  crowded  and  several  optical
components are listed in the WDS.  We add to this list another optical
pair RAO~69FK at 16\farcs4.

HIP~99232 appears  to be  triple. It  was resolved at  0\farcs065 as
WSI~109 (not confirmed yet by a second measure), while Robo-AO finds a
faint companion C at 1\farcs9,  re-observed within a year. We consider
RAO~21AC physical (the  field is not crowded), although  a larger time
base is needed for a solid confirmation because PM(A) is small.

HIP~99572 has a  CPM companion C at 1021\arcsec  ~(AC = TDT~2085) which
was  targeted  by  Robo-AO.  The  18\arcsec ~pair  RAO~88CD  is  
optical (D is not found in 2MASS). 

HIP~99965: we targeted the CPM companion  F ($V=13.5$, AF = GIC~155) at
106\arcsec ~and measured  another star in the field;  the pair RAO~70FL
is optical, its position in 2MASS is very different.

HIP~101234 is an acceleration binary, resolved here for the first time
into a 0\farcs17 pair of equal stars RAO~22, and confirmed with the P200.

HIP~101430 appears to be a  quadruple system. The outer 17\arcsec ~pair
AB=HJ~1535  is composed  of the  spectroscopic and  astrometric binary
Aa,Ab  and a  0\farcs17 pair  Ba,Bb tentatively  seen in  the pipeline
RoboAO image, but not confirmed by speckle processing. This sub-system
is  not listed  in Table~\ref{tab:sec}  for that  reason, but  it was  resolved with the P200  nevertheless.    The  
12\arcsec  ~pair   RAO~71AE  is  optical, with a different position in 2MASS 
in a crowded field.

HIP~102040: The CPM  companion C at 125\arcsec ~from A  (AC = LDS~1045)
was targeted  and resolved at 0\farcs27 (RAO~23Ca,Cb).   The two other
wide companions D and E are  optical; they are not recovered in 2MASS,
while the field is crowded.

HIP~103455  is  an  acceleration  binary  resolved  here  at  0\farcs6
(RAO~24) and confirmed later  with the P200. \citet{MH09} did not resolve
this  system, although  its separation  implies an  orbital  period of
several decades.

HIP~103641 is  a quintuple system  which consists of two  visual pairs
COU~2431Aa,Ab (Aa  is also a close SB)  and HDS~2989Da,Db (HIP~103052)
at 1132\arcsec ~from  A. We targeted D and measured  the close pair, as
well as an  optical star at 19\farcs7 from it  (RAO~72DE), which has a
different  position in 2MASS.   The WDS  components B  and C  are also
optical (crowded field).

HIP~104514  has variable  RV according  to  \citet{Nordstrom2004}.  We
found a new physical companion  at 3\farcs4 and measured it twice. The
companion is confirmed in the 2MASS image and with the P200.

HIP~108473 is  a 7.18-yr  SB with a  new physical companion  RAO~73 at
12\farcs4, confirmed by its position and color in 2MASS.

HIP~109361 is an acceleration binary with variable RV, resolved here
at 0\farcs37 (RAO~26). 

HIP~110291    is   astrometric    and   spectroscopic    binary   with
$P=2.11$~yr.  The  two wide companions B and C measured in  this crowded field
are optical, as evidenced by their  respective positions in 2MASS.

HIP~110574  is  acceleration binary  resolved  here  at 0\farcs09  and
confirmed with the P200 and at SOAR.  The CPM companion  B at 171\arcsec
~($V=15.88$) was not targeted, being too faint.

HIP~110626  is  also an  acceleration  binary,  but  the new  physical
companion  at 4\farcs4  discovered by  Robo-AO (and  confirmed  in the
2MASS image) is  too distant to cause the  acceleration. The system is
therefore triple.

HIP~111148  has a  CPM  companion  B ($V=13.74$)  at  62\arcsec, AB  =
LEP~108. We targeted  B and found two faint stars C  and D at 3\farcs9
and 12\farcs3, respectively. Both companions are considered optical (C
is not  seen in the 2MASS image,  D has a different  position of
268\fdg3, 14\farcs59, and the field is crowded). 

HIP~112935 is triple.  Its CPM  companion D ($V=13.3$, AB = LDS~6388)
at 250\arcsec ~is resolved here  at 0\farcs24 several times at fixed
position during one year, despite the fast PM of 0\farcs5~yr$^{-1}$.

HIP~114456  has a  CPM companion  C at  50\arcsec~\citep{Raghavan2010}
(the WDS pair HJ~1853AB is  optical).  We found only two optical stars
around C, RAO~89CD and CE. They have diffferent positions in 2MASS. 

HIP~115655 has a CPM companion  B = HD~20748 at 185\arcsec. We measure
a faint CD pair at 13\arcsec ~from B. The color of C in 2MASS, $J-K
= 0.65$,  suggests that it is optical, although PM(A) is small and the
position of BC in 2MASS is close to its measured
position.

HIP~116906 has  a CPM companion B at  109\farcs5 (LDS~5112, $V=14.5$),
which  is  resolved  here  into  a 0\farcs5  pair  RAO~31Ba,Bb  (three
measurements within a  year).  The main star hosts  an exo-planet with
$P=572$~d.

HIP~118213 is an acceleration binary. We discover a tertiary companion
at 4\farcs8, confirmed  by its repeated measurement within  a year and
seen in  the 2MASS  image. The  new pair RAO~76  was also  measured with the P200, where the inner binary was resolved as well.

HIP~118225 is an SB  with $P=25.4$~d. We discovered another component
at 5\arcsec,  confirmed by its  repeated measurement and by  the 2MASS
image.

\vspace{0.5cm}
Table~\ref{tab:2MASS}
summarizes the position measurements for the Robo-AO and 2MASS data used to determine if components are physical or optical pairs in many of the newly resolved systems. Some of the Robo-AO pairs were not resolvable by 2MASS, so they are not included in the table.  The table lists the binaries by HIP number, and the components that are being measured, followed by the measured angle (in degrees) and separation (in arcseconds) for the Robo-AO and 2MASS data, and the epoch of observation by 2MASS.  The column $\Delta$ gives the total displacement between the Robo-AO and 2MASS measurements. The next column $\mu \Delta t$ gives the displacement produced by proper motion, assuming $\Delta t$ = 12 years. The last 2 columns are the proper motion of the main target. 

The 2MASS measurements at $\sim$5\arcsec~separations are likely less accurate than its astrometry in general ($\sim$70 mas). The Robo-AO astrometry of wide pairs is accurate to $\sim$0\farcs1~\citep{Law2014a}.  The PM of companion candidates is unknown and comparable to the PM of some targets.  Considering that the relative PM is not known, the assumption of a 12 year time base is good enough for estimating the expected displacement.

\clearpage

\clearpage

\clearpage

\clearpage

\clearpage

\clearpage

\clearpage

\clearpage

\clearpage

\clearpage

\clearpage

\clearpage

\clearpage

\clearpage

\begin{deluxetable}{ll}
\tablewidth{0pt}
\tablecolumns{2}
\tablecaption{Robo-AO Observation Parameters \label{tab:roboao} }
\startdata
\hline 
Observation date range & August 31, 2012 to August 21, 2013   \\
Targets observed & 695 \\
Total observing time &  29 hours \\
Observation filters & SDSS $i'$, LP600  \\
Exposure time & 60 s \\
Typical FWHM resolution & 0\farcs12  to 0\farcs15  \\
Typical contrast ratio at 0.5" &  $\triangle \sim$  5 mag.
\enddata 
\end{deluxetable}

\clearpage

\begin{deluxetable}{llc |ccc|ccc}
\tablewidth{0pt}
\tablecaption{Comparison between speckle and blind deconvolution \label{tab:test} }
\tablehead{
HIP1 & Discoverer  & $N_{\rm obs}$ &\multicolumn{3}{c|}{Speckle (mean, rms)}
&  \multicolumn{3}{c}{Blind (mean, rms)} \\
     & & & $\theta$ & $\rho$ & $\Delta m$ &  $\theta$ & $\rho$ &   $\Delta m$ }
\startdata
20472 & STF~535  & 9 &       271.15 & 1.130 & 0.98 &       271.18 & 1.129 & 1.04 \\
      &          &   & \phn\phn0.32 & 0.002 & 0.10 & \phn\phn0.30 & 0.002 & 0.06 \\
23100 & COU~1525 & 8 &       293.55 & 1.833 & 3.36 &       293.56 & 1.832 & 3.32 \\
      &          &   & \phn\phn0.21 & 0.007 & 0.05 & \phn\phn0.24 & 0.007 & 0.12 
\enddata 
\end{deluxetable}

\clearpage

\begin{deluxetable}{rl c c rrr rrr r}
\tablewidth{0pt}
\tabletypesize{\footnotesize} 
\tablecaption{List of observed components and detection limits  
\label{tab:exprobo_short} }
\tablehead{ 
\colhead{HIP1} &
\colhead{Comp.} &
\colhead{RA (2000)} &
\colhead{Dec (2000)} &
\colhead{$V$} &
\colhead{Sep.} &
\colhead{Epoch} &
\colhead{0\farcs15} &
\colhead{0\farcs8} &
\colhead{2\farcs1} &
\colhead{$\rho_{\rm max}$} \\
 & &   
$^h \; ^m \; ^s$  & 
$^\circ \; ' \; ''$ & 
(mag) &
(\arcsec) & 
+2000 & (mag) & (mag) & (mag) &
(\arcsec)
}
\startdata
   179 &A & 00 02 16.68 & $-$13 24 26.8  &  6.90 &   0.0 &13.5730 &2.7 &5.2 &6.3 & 13.9 \\
   223 &A & 00 02 47.17 & $+$02 07 48.5  &  7.17 &   0.0 &12.7049 &2.1 &4.9 &5.7 & 17.5 \\
   394 &A & 00 04 53.76 & $+$34 39 35.3  &  6.09 &   0.0 &12.7020 &1.7 &5.1 &6.7 & 15.1 \\
   493 &D & 00 05 54.75 & $+$18 14 06.0  &  7.47 &   0.0 &12.6747 &1.2 &4.5 &6.9 & 11.8 \\
   601 &A & 00 07 18.15 & $+$20 57 54.3  &  9.14 &   0.0 &12.7050 &1.8 &4.2 &4.7 & 17.9 \\
   754 &B & 00 09 13.92 & $+$25 16 38.9  & 11.46 &  29.6 &12.6747 &1.6 &4.7 &6.8 & 12.2 \\
\enddata 
 \tablecomments{Table~\ref{tab:exprobo_short} is  published in  its entirety  in  the electronic
 edition; a portion is  shown here for  guidance regarding its  form and content.}
\end{deluxetable}

\clearpage

\begin{deluxetable}{ll@{\,}r@{\,}llccrcrclrrc}   
\tablewidth{0pt}
\tabletypesize{\tiny}       
\tablecaption{Measurements of binary stars with Robo-AO    
\label{tab:double_short} }   
\tablehead{  \colhead{WDS} & \multicolumn{3}{c}{Discoverer} & \colhead{HIP1} & \colhead{Epoch} &  \colhead{N} & \colhead{$\theta$} & \colhead{$\sigma_{\theta}$} &               
\colhead{$\rho$} & \colhead{$\sigma \rho$} & \colhead{$\Delta m$} & \colhead{[O$-$C]$_{\theta}$} & \colhead{[O$-$C]$_{\rho}$} & \colhead{Ref.} \\       
\colhead{(2000)} & \multicolumn{3}{c}{Designation} &  & +2000 &  & \colhead{(deg)} & (deg) & ($''$) & (mas) & (mag) & \colhead{(deg)} & \colhead{($''$)}                   
& \colhead{~} }   
\startdata    
00028$+$0208 & BU  &  281 & AB     & \phn\phn\phn223  & 12.7049 & \phn1 &       160.8 & 0.0 & \phn1.567 & \phn\phn0.5 & 1.66    &         &          &          \\
00057$+$4549 & STT &  547 & AB     & \phn\phn\phn473  & 12.7723 & \phn3 &       186.9 & 0.2 & \phn6.026 &    \phn42.8 & 0.09    &  $-$0.3 &    0.001 & 1 \\ 
             &     &      &        &                  & 13.0478 & \phn2 &       186.4 & 0.1 & \phn5.758 &    \phn68.7 & 0.16    &  $-$0.9 & $-$0.266 & 1 \\ 
00057$+$4549 & POP &  217 & AP     & \phn\phn\phn473  & 13.0478 & \phn2 &       325.5 & 0.3 &    13.140 &    \phn49.4 & 5.03    &         &          &          \\
00073$+$2058 & HDS &   12 &        & \phn\phn\phn601  & 12.7050 & \phn1 & \phn\phn6.7 & 0.1 & \phn1.864 & \phn\phn2.0 & 2.38 :  &         &          &          \\
00251$+$4803 & HDS &   56 & AB     &    \phn\phn1987  & 12.7020 & \phn1 &       279.3 & 0.1 & \phn0.324 & \phn\phn0.5 & 2.19    &  $-$0.8 & $-$0.016 & 2 \\ 
00261$-$1123 & YR  &    4 &        &    \phn\phn2066  & 13.5730 & \phn1 &       176.8 & 0.0 & \phn0.376 & \phn\phn0.0 & 3.12 :  &         &          &          \\
00271$-$0753 & A   &  431 & AB     &    \phn\phn2143  & 13.5730 & \phn1 &       157.6 & 0.1 & \phn0.166 & \phn\phn0.3 & 0.00    &  $-$5.0 & $-$0.019 & 3 \\
\enddata
 \tablecomments{Table~\ref{tab:double_short} is  published in  its entirety  in  the electronic
 edition; a portion is  shown here for  guidance regarding its  form and content.}
\tablerefs{1 - \citet{Kiyaeva2001}, 2 - \citet{Cvetkovic2011}, 3 - \citet{Scardia1981}} 
\end{deluxetable}%
\tablerefs{1 - \citet{Kiyaeva2001}, 2 - \citet{Cvetkovic2011}, 3 - \citet{Scardia1981}, 4 - \citet{Zulevic1997}, 5 - \citet{Cvetkovic2006}, 6 - \citet{Mason2005}, 7 - \citet{Balega2006}, 8 - \citet{Docobo2010}, 9 - \citet{Hartkopf2011b}, 10 - \citet{Soderhjelm1999}, 11 - \citet{Hartkopf1996}, 12 - \citet{Valbousquet1981}, 13 - \citet{Brendley2007}, 14 - \citet{Heintz1996a} , 15 - \citet{Muterspaugh2010b}, 16 - \citet{Starikova1983}, 17 - \citet{Ling2012}, 18 - \citet{Balega2005}, 19 - \citet{Andrade2007}, 20 - \citet{Heintz1998}, 21 - \citet{Scardia2001}, 22 - \citet{Hopmann1958}, 23 - \citet{Hartkopf2009}, 24 - \citet{Hartkopf2010}, 25 - \citet{Starikova1978}, 26 - \citet{Scardia2012}, 27 - \citet{Heintz1963}, 28 - \citet{Alzner1998}, 29 - \citet{Cvetkovic2010}, 30 - \citet{Hartkopf2008}, 31 - \citet{Ling1992}, 32 - \citet{Mason2011}, 33 - \citet{Mason2004a}, 34 - \citet{Hartkopf2000}, 35 - \citet{Zirm2008}, 36 - \citet{Heintz1984}, 37 - \citet{Mason2009}, 38 - \citet{Hartkopf2012}, 39 - \citet{Popovic1969}, 40 - \citet{Docobo2009}, 41 - \citet{Mason2006}, 42 - \citet{Heintz1996b}, 43 - \citet{Heintz1991}, 44 - \citet{Docobo2011}, 45 - \citet{Muterspaugh2010a}, 46 - \citet{Ling2011}, 47 - \citet{Scardia2009}, 48 - \citet{Mason1995}, 49 - \citet{Hale1994}, 50 - \citet{Scardia2007}, 51 - \citet{Mason2012}, 52 - \citet{Alzner2007}, 53 - \citet{Heintz1988}, 54 - \citet{Heintz2001}, 55 - \citet{Mason1999}, 56 - \citet{Seymour2000}, 57 - \citet{Heintz1997}, 58 - \citet{Zirm2011}, 59 - \citet{Heintz1975}, 60 - \citet{Heintz1986a}, 61 - \citet{Heintz1994}, 62 - \citet{Docobo2008}, 63 - \citet{Eggenberger2008}, 64 - \citet{Hartkopf1989}, 65 - \citet{Aristidi1999}, 66 - \citet{Scardia2003}, 67 - \citet{Zeller1965}, 68 - \citet{Scardia2008}, 69 - \citet{Muterspaugh2008}, 70 - \citet{Fekel1997}, 71 - \citet{Heintz1995}, 72 - \citet{Heintz1986b}, 73 - \citet{Prieur2010}, 74 - \citet{Griffin1987}, 75 - \citet{Mason2004b}, 76 - \citet{Seymour2002}, 77 - \citet{Hartkopf2011a}}

\clearpage

\end{document}